\documentclass{aa}  

\usepackage{graphicx}
\usepackage[dvipsnames]{xcolor}
\usepackage{txfonts}
\usepackage{hyperref}
\begin{document}

   \title{Advancing accuracy in age determinations of old-disk stars using an oscillating red giant in an eclipsing binary}
   \titlerunning{Accuracy of ages for evolved old stars}

   \author{J. S. Thomsen\inst{1,2,3}
   \and A. Miglio\inst{1,2,4}
   \and K. Brogaard\inst{3,1}
   \and J. Montalbán\inst{1,4}
   \and M. Tailo\inst{5,1}
   \and W. E. van Rossem\inst{1}
   \and G. Casali\inst{6,7,1,2}
   \and D. Jones\inst{8,9,10}
   \and T. Arentoft\inst{3}
   \and L. Casagrande\inst{6,7}
   \and D. Sebastian\inst{4}
   \and G. Buldgen\inst{11}
   \and A. Triaud\inst{4}
   \and M. Matteuzzi\inst{1,2}
   \and A. Stokholm\inst{4,1,3}
   \and M. N. Lund\inst{3}
   \and B. Mosser\inst{12}
   \and P. F. L. Maxted\inst{13}
   \and J. Southworth\inst{13}
   \and J. T. Gadeberg\inst{10}
   \and N. Koivisto\inst{14,10}
   \and Z. Gray\inst{15,16,10}
   \and V. Pinter\inst{10,17,18}
   \and K. Matilainen\inst{14,10}
   \and A. A. Djupvik\inst{10}
   \and J. Jessen-Hansen\inst{3}
   \and F. Grundahl\inst{3}
   \and D. Slumstrup\inst{19,8,20}
   \and S. Frandsen\inst{3}
   }
    \institute{
    Dipartimento di Fisica e Astronomia, Università di Bologna, Via Zamboni 33, Bologna, 40126, Italia\\
    \email{jet@phys.au.dk}
    \and
    Osservatorio di Astrofisica e Scienza dello Spazio di Bologna, INAF, Via Gobetti 93/3, Bologna, 40129, Italia
    \and
    Stellar Astrophysics Centre, Department of Physics and Astronomy, Aarhus University, Ny Munkegade 120, Aarhus C, 8000, Denmark
    \and
    School of Physics and Astronomy, University of Birmingham, B15 2TT, Birmingham, United Kingdom
    \and
    Osservatorio Astronomico di Padova, INAF, Vicolo dell'Osservatorio 5, Padova, 35122, Italia
    \and
    Research School of Astronomy \& Astrophysics, Australian National University, Cotter Rd., Weston, ACT 2611, Australia
    \and
    ARC Centre of Excellence for All Sky Astrophysics in 3 Dimensions (ASTRO 3D), Stromlo, Australia
    \and
    Instituto de Astrof\'isica de Canarias, E-38205, La Laguna, Tenerife, Spain
    \and
    Departamento de Astrof\'isica, Universidad de La Laguna, E-38206, La Laguna, Tenerife
    \and
    Nordic Optical Telescope, Rambla Jos\'e Ana Fern\'andez P\'erez 7, Bre\~na Baja, La Palma, 38711, Spain
    \and
    Institut d'Astrophysique et G\'eophysique, l'Universit\'e de Li\`ege, ll\'ee du 6 ao\^ut 17, Li\`ege, 4000, Belgium
    \and
    LIRA, Observatoire de Paris, Universit\'e PSL, Sorbonne Universit\'e, Universit\'e Paris Cit\'e, CY Cergy Paris Universit\'e, CNRS, 92190 Meudon, France
    \and
    Astrophysics Group, Keele University, Staffordshire, ST5 5BG, United Kingdom
    \and
    Department of Physics and Astronomy, University of Turku, Turku, 20014, Finland
    \and
    Department of Physics, P.O. Box 64, FI-00014 University of Helsinki, Finland.
    \and
    Armagh Observatory and Planetarium, College Hill, BT61 9DG, Northern Ireland, UK
    \and
    University of Craiova, Alexandru Ioan Cuza 13, Craiova, 200585, Romania
    \and
    CAHA, Observatorio de Calar Alto, Sierra de los Filabres, Gérgal, 04550, Spain
    \and
    GRANTECAN, Cuesta de San José s/n, E-38712, Breña Baja, La Palma, ES
    \and
    European Southern Observatory, Alonso de Cordova 3107, Vitacura, Santiago, Chile
    }

   \date{Received XX; accepted YY}

 
  \abstract
   {The study of resonant oscillation modes in low-mass red giant branch stars enables their ages to be inferred with exceptional ($\sim 10\%$) precision, unlocking the possibility to reconstruct the temporal evolution of the Milky Way at early cosmic times. Ensuring the accuracy of such a precise age scale is a fundamental yet difficult challenge. Since the age of red giant branch stars primarily hinges on their mass, an independent mass determination for an oscillating red giant star provides the means for such assessment.}
   {We analyze the old eclipsing binary KIC\,10001167, which hosts an oscillating red giant branch star and is a member of the thick disk of the Milky Way. Of the known red giants in eclipsing binaries, this is the only member of the thick disk that has asteroseismic signal of high enough quality to test the seismic mass inference at the $ 2\%$ level.} 
   {We measure the binary orbit and obtain fundamental stellar parameters through combined analysis of light curve eclipses and radial velocities, and perform a detailed asteroseismic, photospheric, and Galactic kinematic characterization of the red giant and binary system.}
   {We show that the dynamically determined mass $0.9337\pm0.0077 \rm\ M_{\odot}$ (0.8\%) of this 10 Gyr-old star agrees within $1.4\%$ with the mass inferred from detailed modelling of individual pulsation mode frequencies (1.6\%). This is now the only thick disk stellar system, hosting a red giant, where the mass has been determined both asteroseismically with better than $2$\% precision, and through a model-independent method at $1$\% precision, and we hereby affirm the potential of asteroseismology to define an accurate age scale for ancient stars to trace the Milky Way assembly history.} 
   {}


   \maketitle
%

\section{Introduction}

Precise age-dating of cosmic structures is one of the key challenges of modern astrophysics.
The availability of chemo-dynamical constraints for millions of stars, thanks to the \emph{Gaia} mission \citep{Gaia2016} and large-scale spectroscopic surveys, signified a step change in our understanding and identification of stellar populations that constitute the Milky Way \citep[e.g., see][]{Helmi2020, Belokurov2022,  Queiroz2023, RecioBlanco2023, Gallart2024}. 
Moreover, information about disk galaxies at high redshift (z) are becoming available thanks to observations with JWST and ALMA \citep[e.g., see][]{Ferreira2023, Roman-Oliveira2023, Tsukui2021} and we now have the possibility of comparing the high-z picture of galaxies with that given by the oldest of stars within our Galaxy for which we have exquisitely high-resolution information about their dynamical and chemical composition.
To chronologically connect these complementary views we need a high ($\sim10\%$) temporal resolution, especially for the oldest tracers, i.e. stars with ages of $\sim$10 Gyr.
Precise and accurate ages of the oldest objects in the Universe would also constitute a crucial test for modern cosmology \citep[e.g., see][]{Cimatti2023}.

A significant step forward in the challenging task of inferring precise and accurate stellar ages \citep{Soderblom2010} has been provided by asteroseismology, enabling direct information about stellar interiors to become accessible for scientific investigation.
Evolved low-mass stars showing solar-like oscillations, due to both their intrinsic brightness and long main sequence (MS) lifetime, represent ideal clocks to infer the chronology of structure formation in the Milky Way, allowing precise age-dating of the oldest objects in the Galaxy \citep{Chaplin2020, montalban2021, Borre2022}.
Ages inferred using seismic constraints are often adopted as training sets to extend the age inference to hundreds of thousands of stars using, e.g.,  machine learning techniques applied to stellar spectroscopy \citep[e.g., see][and references therein]{Anders2023}.
Independent verification of the asteroseismic age scale is thus of paramount importance, since solar-like oscillating giants are starting to take on the role of primary calibrators to what is effectively becoming the cornerstone of the stellar age scale.

Since the age of red giant branch (RGB) stars is effectively their MS lifetime, which primarily depends on the initial mass (age $\propto M^{-\alpha}$, with $\alpha \sim 3$ \citep[e.g., see][]{Kippenhahn2013}), the reliability of the age scale of these stars is anchored to our ability to accurately measure their masses.
In the last decade, several efforts were dedicated to comparing asteroseismically inferred masses with independent determination of masses available for stars in clusters \citep[see][and references therein]{Handberg2017, Brogaard2021Hyades, Brogaard2023}, and eclipsing binaries \citep[see][and references therein]{Gaulme2016,Brogaard2018,theme2018,thomsen2022, brogaard2022}.
While these independent measurements offer valuable tests, their precision or exploration of the parameter space are somewhat limited, which poses a significant obstacle especially for old, low-mass, metal-poor giants, as independent and precise measurements are rare in this part of the parameter space. For example, constraints from globular clusters are currently limited to a few giants with detectable oscillations using K2 data \citep[e.g.][]{Tailo2022, Howell2024} of significantly lower asteroseismic quality than \emph{Kepler}, while the only other confirmed \emph{Kepler} thick disk eclipsing binary, KIC\,4054905 \citep{Gaulme2016, brogaard2022} has significant light contamination and lower oscillation amplitudes, in both cases effectively limiting the asteroseismic precision such that it cannot challenge the mass accuracy at the 2\% precision level typically achievable using individual mode frequencies from long-duration observations \citep{montalban2021}.
A fundamental challenge is thus to obtain a model-independent mass determination with percent-level precision for an old, metal-poor red giant star showing well-determined solar-like oscillations.

In this context, the detached, long-period eclipsing binary KIC\,10001167 bears the hallmarks of the ideal benchmark for the mass and age scale of old stars.
This system is bright (\emph{G}$=10.05$) and was observed for 4 years by the \emph{Kepler} space satellite \citep{Borucki2010}, providing an exquisite photometric monitoring, a detailed characterisation of the eclipses of both its red giant (RG) and MS component, as well as the detection of solar-like oscillations in its RG star.
Moreover, KIC\,10001167 is reported to be a spectroscopic double-lined binary in \citet{Gaulme2016}, thus enabling a model-independent inference of the stellar component masses.

In \citet{Gaulme2016}, however, they obtained a dynamical mass of $0.81\pm 0.05\,\rm M_{\odot}$ ($6.2\%$ precision) for the red-giant component. Such a low value is puzzling as it would imply an age older than the currently accepted age of the Universe, as already noted in \citet[][]{Brogaard2018}. The low radial velocity (RV) precision of the \citet{Gaulme2016} study has been explored in detail in previous works \citep{Brogaard2018, thomsen2022, brogaard2022}, and the poor sampling of the RV semi-amplitudes for this particular system has likely exacerbated this limitation. Moreover, an asteroseismic study of the individual oscillation modes of the star reported a mass of $0.94\pm 0.02\,\rm M_{\odot}$ \citep{montalban2021}, significantly higher and of higher precision ($2.1$\%) compared to the dynamical measurement.

To investigate these potential discrepancies and limitations, we present revised dynamical mass measurements of the system components based on long-term high precision RV monitoring, together with an in-depth spectroscopic, photometric, and asteroseismic analysis and modelling of the RG in the system.

\section{Methods}\label{sec:methods}
The light curve photometry used in this work is from the \emph{Kepler} space mission \citep{Borucki2010}. For spectroscopic characterization and RV sampling of the binary orbit, we have obtained 45 spectroscopic follow-up observations with the Fibre-fred Echelle Spectrograph (FIES) \citep{Telting2014} at the Nordic Optical Telescope (NOT) on La Palma, having spectral resolution $R\sim67.000$. Spectral extraction and wavelength calibration is performed by the \emph{FIEStool} \citep{FIEStool} observatory pipeline.

The following sections describe the various methods employed in our analysis. The RV measurement and spectral component separation is presented in Sect.~\ref{sec:rv-sep}. Photospheric analysis of the RG through spectroscopy and photometry is presented in Sect.~\ref{sec:specanal} and \ref{sec:phot}, respectively. Sect.~\ref{sec:binary} demonstrates the combined analyses of eclipse photometry and radial velocities. Sect.~\ref{sec:kinematic} outlines the Galactic kinematic analysis. Sect.~\ref{sec:asteroseismology} details the methods to obtain observational asteroseismic constraints for the RG. Finally, Sect.~\ref{sec:inference} illustrates how stellar parameters are inferred through comparison with stellar models.

\subsection{Radial velocity and spectral component separation}\label{sec:rv-sep}
For simultaneous RV measurements and spectral separation, we use the Python code \href{https://jsinkbaek.github.io/sb2sep/}{\emph{sb2sep}} (v. 1.2.15) from \citet{thomsen2022}.
It employs the broadening function formulation \citep[][]{Rucinski2002} with synthetic templates from \citet{Coelho2005} for RVs, and the spectral separation method of \citet{Gonzalez2006}.
To reduce instrumental drift, the wavelength solution is defined using a Thorium-Argon (ThAr) spectrum captured immediately before observation, and telluric RV corrections are applied.
Barycentric corrections and barycentric Julian dates are calculated using \emph{barycorrpy} \citep{kanodia2018}.
Outputs from the separation and RV extraction are reported in Appendix Table~\ref{table:rvs}, while further details can be found in Appendix~\ref{sec:rv_details}. With a completely independent analysis method, outlined in Appendix~\ref{sect:rv-residual}, we find consistent RV variation.

Excluding (including) the jitter term we determine in Appendix~\ref{sec:rv_details} and included in the binary orbit fit, the mean RV uncertainty is $29$ ($96$) m/s for the RG and $0.48$ ($0.49$) km/s for the MS star.

By separating the stellar components of the spectra, we can obtain a high Signal-to-Noise ratio (SNR) stacked spectrum of the RG ideal for spectroscopic analysis. The separated component spectrum of the RG has nominal SNR of $\sim 270$ for the RG.
The MS component spectrum is dominated by noise from the RG and of insufficient quality for atmospheric analysis.
\subsection{Spectral Analysis}\label{sec:specanal}
A detailed review on spectral analysis methods and their wide applications can be found in \citet{Nissen2018}.

The system has been observed as part of the intermediate-resolution ($R\sim 22.500$) near-infrared spectroscopic survey APOGEE DR17 \citep{Apogeedr17}. Since we have obtained a high-resolution, high-SNR component spectrum of the RG from the optical FIES spectra, we can perform an independent characterization. The separated spectrum was renormalized using a wavelength dependent light ratio derived from the \emph{Kepler} passband light ratio obtained from eclipsing binary analysis in Sect.~\ref{sec:binary}, assuming a black body spectral energy distribution, which is sufficient since the luminosity ratio is low ($\rm \frac{L_{MS}}{L_{RG}}\sim 1.8\%$).

Atmospheric parameters were then determined from classical equivalent width (EW) measurements obtained with DOOp \citep[Daospec Output Optimiser pipeline,][]{cantat2014}, a pipeline wrapper of DAOSPEC \citep{stetson08}. DAOSPEC is a fortran program for automatic recovery and identification of stellar absorption lines from an input line-list, continuum fitting, and EW measurement. To derive atmospheric parameters from EWs, we used FAMA \citep[Fast Automatic MOOG Analysis,][]{magrini13}, which is an automated version of MOOG version 2017 \citep{sneden12}, a one-dimensional local-thermal-equilibrium radiative transfer code that can be used to derive abundances from EWs through spectral synthesis. FAMA uses MOOG together with MARCS model atmospheres \citep{Gustafsson2008}.
We fix $\log g$ to the value inferred using asteroseismic constraints, and determine the other atmospheric parameters through the excitation equilibrium, by minimization of the trend between the reduced EW, $\log(EW/\lambda)$.
FAMA computes elemental abundances using the MOOG routines {\sc abfind} and {\sc blends}, see \citet{magrini13} for further details.

We use the line-list given in \citet{Slumstrup2019}, which is curated to avoid saturated lines, having only lines with EW $< 80$ m\AA. It also includes astrophysically calibrated oscillator strengths. We compare astrophysical and laboratory oscillator strengths in Appendix~\ref{sect:linelist} to validate our choice. We adopt a total uncertainty of $0.1$dex on $\rm [Fe/H]$ and $\rm [\alpha/Fe]$ following the investigations of \citet{Bruntt2010} (see Table~\ref{table:abundances} for statistical uncertainty).

We obtain elemental abundance measurements of neutral atomic lines NaI, MgI, AlI, SiI, CaI, TiI, CrI, FeI, NiI, and singly-ionized lines TiII, FeII, as well as the logarithmic abundance of alpha-process elements, [$\alpha$/Fe], here defined as $\frac{1}{4}\left( [\rm Ca/Fe] + [\rm Si/Fe] + [\rm Mg/Fe] + [\rm Ti/Fe]\right)$. The solar abundances used for our analysis are those from \citet{Asplund2009}, while APOGEE DR17 used \citet{grevesse2007}. The elements are recorded in logarithmic abundance relative to iron in Table~\ref{table:abundances}, with only statistical uncertainty. We note that the agreement with APOGEE DR17 on [Fe/H] is 0.05dex, and in fact better than $\lesssim 0.01$ dex if the difference in solar scale is accounted for.
\subsection{Eclipsing binary analysis}\label{sec:binary}
Analysis of spectroscopic double-lined binaries showing eclipses is a fundamental method of measuring precise and accurate stellar masses and radii. For an observational review see, e.g., \citet{Torres2010}, while a detailed theoretical background on the physics of advanced eclipsing binary modelling is available in \citet{Prsa2018}.
We perform two independent combined eclipsing light curve and RV analyses, using the codes JKTEBOP \citep[v. 43,][]{Southworth2013} and PHysics Of Eclipsing BinariEs 2 \citep[PHOEBE 2][]{conroy2020}. Properties of KIC\,10001167 determined by our eclipsing binary analyses are presented in Table~\ref{table:EBdata}. JKTEBOP is an eclipsing binary fitting code which offers very high computational efficiency and numerical precision through a few key analytic approximations. Particularly, during eclipse the two stellar components are assumed to be perfectly spherical, while during out-of-eclipse modelling they can be treated as either spherical or bi-axial ellipsoids. PHOEBE 2 instead offers the possibility of relaxing several of these analytic approximations, thereby achieving higher accuracy for stars with significant deformation and reflection, at the cost of considerably lower computational efficiency. One of the main such features is the numerical approximation of the surface of the stars as a discretized mesh of connected triangles deformable by a Roche lobe potential following \citet{Wilson1979}. It also includes internal handling of limb darkening, derived from model atmospheres for each mesh-point, unlike JKTEBOP where an analytic prescription must be assumed.
\subsubsection{JKTEBOP}\label{sec:jktebop}
For eclipsing binary analysis with JKTEBOP, we use the \emph{Kepler} mission \citep{Borucki2010} Presearch Data Conditioning light curve \citep[PDCSAP, ][and references therein]{PDCSAP_3_2017}\footnote{\url{https://mast.stsci.edu/portal/Mashup/Clients/Mast/Portal.html}\label{footnote:mast}}. The choice of light curve is explained further in Appendix~\ref{sec:jktebop-details}, as well as the pre-processing we perform by normalizing the eclipses with polynomial fitting since we will be treating the stars as spherical during JKTEBOP analysis.

We find no evidence for background contamination from nearby stars (see Appendix~\ref{sec:jktebop-details}), nor indications of significant in-system contamination from the spectroscopy (see Appendix~\ref{sec:spec-tlight}).

We apply the ($\rm h_1$, $\rm h_2$) parameterized power-2 limb darkening law with coefficients interpolated from \citet{Claret2022}, as the ($\rm h_1$, $\rm h_2$) parametrization has been found to be superior to other two-parameter prescriptions when fitting for one coefficient \citep{maxted2023}. Further details can be found in Appendix~\ref{sec:eb-appendix}, particularly \ref{sec:jktebop-details}, as well as \ref{sec:limbdark} and \ref{sec:eb-atmos} from which we estimate a systematic uncertainty of $\sim 0.7\%$ for the radius of the RG from limb darkening and atmosphere approximations.

An evaluation of the light curve residuals around the best-fit demonstrates that they are dominated by stochastic solar-like oscillations, rather than statistical noise. 
Therefore, as in \citet{thomsen2022}, a residual block bootstrap resampling method of the light curve is employed for uncertainty estimation.
For the radial velocities, the sampling method also includes Monte Carlo simulation on top of residual resampling, a new addition since \citet{thomsen2022}.

Root mean square (RMS) of the residuals of our RVs from FIES is $0.097$\,km/s for the giant and $0.42$\,km/s for the MS component.
There is a clear residual signal in the RVs of the RG after subtracting the binary RV curve, which we investigate in Appendix~\ref{sect:rv-residual}. Despite this extra signal, the SNR limited precision of $0.42$ km/s for the MS star RVs still dominates the stellar mass error budget.

In Appendix~\ref{sec:ltte}, we investigate and conclude that light travel time, while significant, is not necessary to account for to obtain accurate stellar parameters for this system.
\subsubsection{PHOEBE 2}\label{sec:phoebe}
For the analysis with PHOEBE 2, we perform a custom, iterative filtering of the KASOC light curve, inspired by \citet{Handberg2014}, in order to keep the full eclipsing binary signal. An explanation on the choice of light curve, as well as a description of the filtering, can be found in Appendix~\ref{sec:phoebe-details}.

Then, we perform affine-invariant Markov chain Monte Carlo sampling (MCMC) with \emph{emcee} \citep{ForemanMackey2013}, starting from the best fit JKTEBOP solution.
We find it necessary to heavily bin the data in order to reduce computing time.
We bin the data in time-space, with 2-day binning outside of eclipses, no binning during eclipse ingress/egress, and 0.3 day binning within total and annular eclipse.

There is clear evidence of Doppler boosting/beaming in the light curve. While PHOEBE 2 does not officially support boosting in the current version due to numerical issues with its native interpolation of coefficients, we manually re-enable user-provided boosting coefficients to be supplied, allowing us to sample it as a free parameter. This functionality will be made available in the next feature release v2.5 (Jones et al. in prep).
As a result of our sampling choice, $T_{\rm eff, RG}$ is poorly constrained for this analysis since the boosting coefficient is completely uncoupled.

Our uncertainties obtained from the PHOEBE MCMC sampling are heavily underestimated due to the presence of correlated (asteroseismic) un-modeled signal in the data (see Appendix~\ref{sec:phoebe-details}). Our JKTEBOP uncertainties should therefore be used instead when comparing with other analyses, and we refer to the JKTEBOP result when comparing with asteroseismic, and photometric, inference.

We remark that the agreement on essential light curve fit parameters between the two methods is $0.4\sigma$ for the sum of the fractional radii $r_{\rm MS}+r_{\rm RG} = \frac{R_{\rm MS}+R_{\rm RG}}{a}$, $0.2\sigma$ for the radius ratio $k = \frac{R_{\rm MS}}{R_{\rm RG}}$, and $0.2\sigma$ for the inclination $i$ (assuming JKTEBOP uncertainties). These are the free parameters expected to be most significantly affected by the treatment of the stellar surface shape. For reference, the agreement on the radius of the giant is $0.4\sigma$. This indicates that a spherical treatment of the stars during eclipse is sufficient for accurate analysis of this system, provided that proper pre-processing of the light curve is performed.
\subsection{Parallax, photometry, and IRFM}\label{sec:phot}
Gaia DR3 \citep{Gaia2016, gaia2022} offers parallax and optical photometry for KIC\,10001167, while 2MASS \citep{2MASS2006} provides near-infrared photometry. We present here the main steps of our photometric analysis, while further details can be found in Appendix~\ref{sec:parallax-phot-irfm}.
Table~\ref{table:astrometry} shows the astrometric parameters from \emph{Gaia} DR3, including an additional uncertainty estimate due to the potential effect of the binary orbit, which we derive in Appendix~\ref{sec:parallax-error}.

A detailed description of the infra-red flux method (IRFM) can be found in \citet{casagrande2006}, but we summarize the principles here. Given a set of photometric observations covering a wide wavelength range, in this work \emph{Gaia} DR3 BP, G, RP, and 2MASS J, H, Ks, the majority of the bolometric flux of the star can be measured directly. The remainder (typically $15-30$\%) is predicted with model fluxes (in this work from \citealt{ck03}) to produce a bolometric correction assuming an initial effective temperature $T_{\rm eff}$. 
For a star of given angular size $\theta$, the IRFM exploits that the bolometric flux has a strong sensitivity to $T_{\rm eff}$ through the Stefan-Boltzmann law, while the infrared flux has a linear dependence on $T_{\rm eff}$ through the Rayleigh-Jeans curve for stars hotter than about 4000 K. The ratio of bolometric to infrared flux is used to eliminate the dependence on $\theta$, while preserving a good sensitivity to $T_{\rm eff}$ \citep[see e.g. Fig. 1 in][for an illustration]{Blackwell1979}. A new $T_{\rm eff}$ can thus be obtained, which is then iterated until convergence is reached in temperature. Since both $T_{\rm eff}$ and bolometric flux are determined at each iteration, $\theta$ can also be computed.

Table~\ref{table:phot-rg} includes photometric measurements of the RG derived using the parallax and the IRFM measurements with the implementation described in \citet{casagrande2021}, as well as single-passband luminosity estimates involving bolometric corrections \citep[][and references therein]{casagrande2018b}. While the IRFM is known to be nearly model independent and only mildly affected by the adopted metallicity and surface gravity \citep[e.g.][and references therein]{casagrande2006}, it critically depends on the input photometry and reddening. For this analysis, we have accounted for the presence of both stellar components in the photometry, using the radius ratio (JKTEBOP) and effective temperature ratio (PHOEBE) obtained from the eclipsing binary analysis. The reddening for this system is small, and does not affect the results at the agreement level of available dust maps, which we demonstrate in Appendix~\ref{sec:extinction},~\ref{sec:irfm}. More details can be found in Appendix~\ref{sec:irfm}.

\subsection{Kinematics}\label{sec:kinematic}
We measure the Galactic position and velocities of KIC\,10001167 using the distance derived from the parallax (see Appendix~\ref{sec:distance}), the celestial position and proper motions from \emph{Gaia} DR3, and system RV from the JKTEBOP RG RV solution.

The Galactic orbital kinematics and the integrals of motion of the star are calculated using the \textsc{galpy} fast orbit estimation algorithm \citep{Bovy2015,Mackereth2018} by adopting the \texttt{McMillan2017} potential \citep{mcmillan2017}. For the Sun and the Local Standard of Rest, we assume that the Sun's distance to the Galactic centre is $R_{\odot} = 8.2$ kpc \citep{mcmillan2017} and that the solar movement is $(U_{\odot}, V_{\odot}, W_{\odot}) = (11.1, 12.24, 7.25$) km~s$^{-1}$  \citep{Schonrich2010} with v$_\mathrm{LSR}$ = $221$~km~s$^{-1}$.
The uncertainties on the dynamical quantities are calculated using a bootstrap method, which involves randomly selecting a sample of phase-space quantities based on the given observational uncertainties and the covariance matrix associated with the \emph{Gaia} parameters. 

In Table~\ref{table:astrometry}, astrometric and kinematic measurements are given. 
\subsection{Asteroseismic constraints}\label{sec:asteroseismology}
To measure the properties of KIC1000167's  solar-like oscillation spectrum, we use both the \emph{KEPSEISMIC} \citep[e.g.][and references therein]{Pires2015} (From MAST\footref{footnote:mast}) and the KASOC photometric light curve, both designed for asteroseismic analysis of giants. The KASOC light curve pipeline employs a filtering technique made to remove transit signals (see \citet{Handberg2014} for details), while the KEPSEISMIC light curve version we use is filtered with an 80-day window. 
\subsubsection{Individual mode oscillation frequencies}\label{sec:freq_numax}
We measure individual oscillation frequencies using four different combinations of pipelines and light curve reductions.
The methods are labelled with the power spectrum name (KEPSEISMIC or KASOC), and by the background and frequency extraction pipeline. The latter are either FAMED \citep[][and references therein]{corsaro2020}, PBJam \citep{nielsen2021}, or the frequency extraction method described in \citet{Arentoft17}. As background models describing stellar granulation, activity and white noise we use either a set of three Harvey-like profiles or the model described in \citet{Arentoft17} (see Table~\ref{table:RGseis}).

Our frequency measurements are detailed in Appendix~\ref{sec:asteroseismology_appendix}, and further in \ref{sec:frequency_extraction}. 
Frequencies measured using different methods agree within $\sim 1\sigma$ except for a few of those recovered by KEPSEISMIC+FAMED. We define the reference set of frequencies to use in the modelling as those recovered with PBJam.
During the later model inference, we compare with an inference performed using the set of frequencies showing the largest difference from PBJam (KEPSEISMIC+FAMED), and treat it as a systematic source of uncertainty in the recovered stellar parameters.

Before use for asteroseismic inference, the observed frequencies are corrected for the Doppler shift produced by the system line-of-sight velocity following \citet{davies2014}. However, due to the low pulsational frequencies of the modes, the shift is negligible in comparison to the frequency uncertainty.
\subsubsection{Average asteroseismic parameters}\label{sec:avg_ast}
Table~\ref{table:RGseis} shows the measured global asteroseismic parameters average large frequency separation ($\Delta\nu$) and frequency of maximum oscillation power ($\nu_{\rm max}$), along with literature results for $\nu_{\rm max}$.
To determine $\Delta\nu$, we use first a power spectrum stacking method ($\Delta\nu_{\rm PS}$) and then refine it using only the individual radial-mode oscillation frequencies ($\Delta\nu_{0}$) \citep{Arentoft17, Brogaard2021Hyades}.
$\nu_{\rm max}$ is obtained using the methods mentioned in Sect.~\ref{sec:freq_numax}. For the stellar parameter inference described below, a conservative estimate of $\nu_{\rm max}=19.93\pm 0.47 \mu$Hz is adopted, which keeps all the measurements in Table~\ref{table:RGseis} within 1$\sigma$. 
\subsection{Stellar parameter inference}\label{sec:inference}
Stellar parameters are inferred by comparing seismic and non-seismic observational constraints with predictions from models of stellar structure and evolution. We employ two grids of models based on different stellar evolution and pulsation codes. 

The first grid was already presented in \citet{montalban2021} and is based on the Liège stellar evolution code CL\'ES \citep{scuflaire2008a}. Stellar models are evolved from the pre-main sequence up to a radius of 25 $\rm R_{\odot}$  on the RGB.  Adiabatic oscillations of radial modes are computed with the code LOSC \citep{scuflaire2008b}. The grid, together with a description of its key input physics, is available at \url{https://zenodo.org/records/4032320}.

The second stellar models grid is described in detail in Tailo et al. (in prep) and is computed using the stellar evolution software \textsc{MESA} \citep[][and references therein]{paxton2019} in its version n. 11701. Details on the grid can also be found in Appendix~\ref{sec:mesa}.

We use the code Asteroseismic Inference on a Massive Scale \citep[AIMS][]{Rendle2019} to infer stellar parameters, and to explore the impact on the estimated mass and radius of using different combinations of observational constraints and uncertainties in the modelling. AIMS is a Bayesian parameter inference code, which provides best-fitting stellar properties and full posterior probability distributions by comparing observational constraints with theoretical predictions from stellar models. It samples the parameter space using MCMC, and includes interpolation routines capable of handling multi-dimensional irregularly sampled stellar model grids. For our asteroseismic inference, we supply AIMS with observed individual oscillation frequencies extracted from the power spectrum, we constrain the frequency of maximum power (see Sect.~\ref{sec:avg_ast}), and we include an observational constraint on the effective temperature and surface metallicity $\rm (Z/X)_{surf}$.
Using individual frequencies as observational constraints contributes to significantly reducing the uncertainties affecting estimated global stellar parameters. As demonstrated in several studies \citep[e.g., see][]{Gough1990}, theoretical individual-mode frequencies should be corrected for the so-called surface effects, i.e. systematic uncertainties stemming from our limited ability to model the star's near-surface layers. In this study we correct the theoretical frequencies using a two-term prescription following \citet[Eq. 4]{ballgizon2014} that involves two free parameters to be derived by the fitting procedure, a cubic $a_3$ and an inverse term $a_{-1}$ such that the correction $\delta\nu$ becomes
\begin{align}
    \delta\nu(\nu, \mathcal{I}) = \left[a_{-1}(\nu/\nu_{\rm ac})^{-1}+a_3(\nu/\nu_{\rm ac})^3\right]/\mathcal{I},
\end{align}

\noindent where $\nu$ is the theoretical mode frequency, $\nu_{\rm ac}$ is the acoustic cut-off frequency, and $\mathcal{I}$ is the normalized mode inertia.
The other free parameters that are sampled are the stellar mass, the initial mass fraction of metals and the stellar age. All other stellar parameters are derived from these or held constant (when indicated), except for one fit where we explore the effect of varying the initial helium fraction $Y_i$ as well.

The main, reference fit is obtained using the CL\'ES grid ($[\rm \alpha/Fe] = 0.3$), with observational constraints from the 6 radial mode frequencies shown in Fig.~\ref{fig:article_psd} (obtained with PBJam), the quoted $\nu_{\rm max}$ value from Sect.~\ref{sec:avg_ast}, as well as metallicity and effective temperature from APOGEE DR17. In Appendix~\ref{sec:inference-details}, we explore the effect of the choice of observational input in detail through several runs of AIMS and use that to infer realistic systematic uncertainties.

Finally, we also use AIMS to infer the age including  as observational constraints the dynamical mass and radius of the RG instead of the oscillation frequencies. To estimate systematic uncertainties, we follow the same treatment as highlighted in the previous paragraph for the inferences using oscillation frequencies (variation in effective temperature and metallicity source, change to grid [$\alpha$/Fe], choice of stellar evolution code).

We also provide asteroseismic scaling relation measurements in Appendix~\ref{sec:scaling-relations}, but stress that scaling relations can be systematically much more uncertain than individual frequency inferences, which is why we do not put focus on these results in the paper.
%
%
%
\section{Results}
In this section we summarize the analysis results for KIC\,10001167.
\subsection{Spectroscopic, photometric, and kinematic analysis}\label{article:target}
\begin{figure}
    \centering
    \includegraphics[width=\columnwidth]{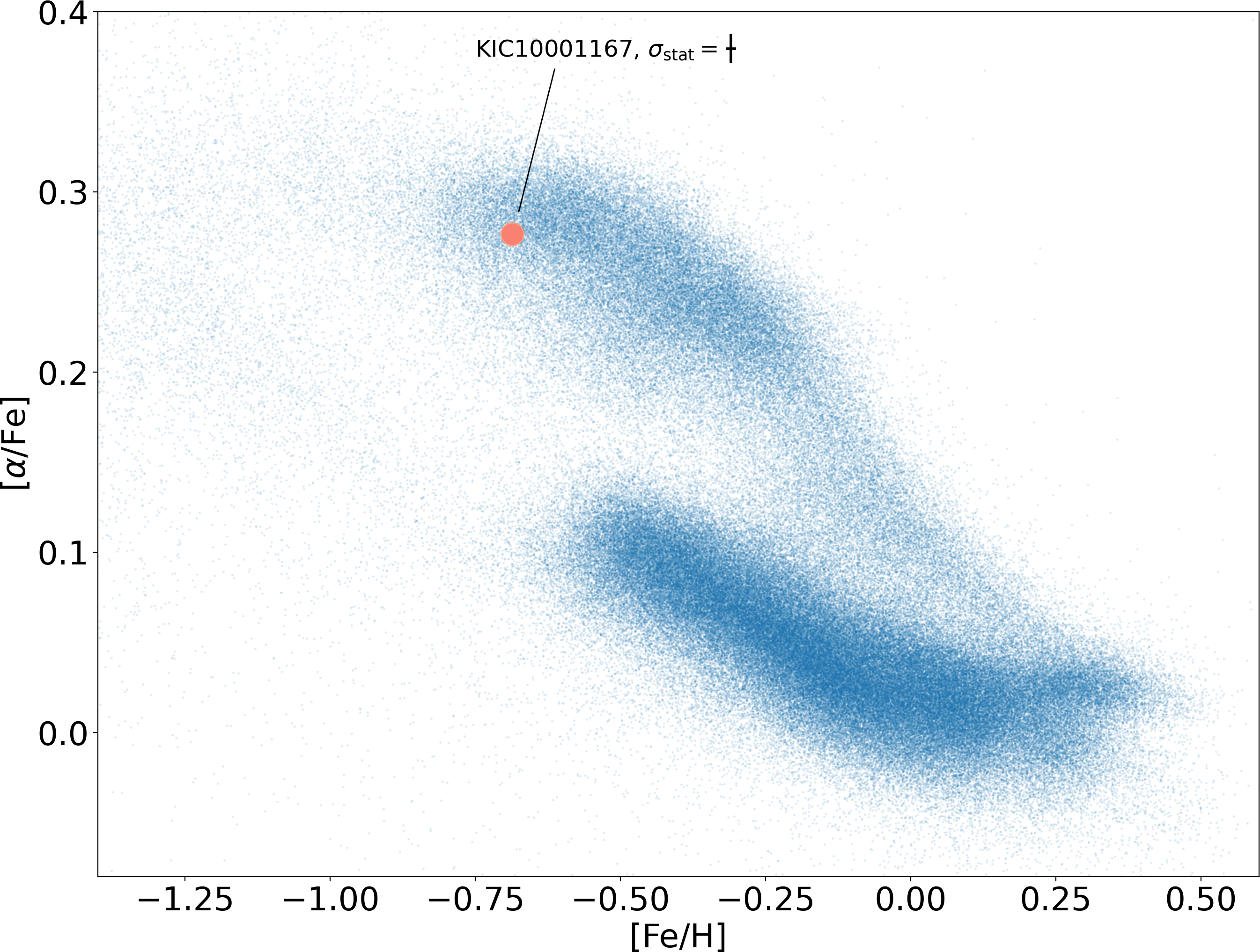}
    \caption{$\alpha$-enhancement level vs. iron-abundance from APOGEE DR17, for stars with $1.5 < \log{g} < 3$, with KIC\,10001167 highlighted.}
    \label{fig:article_alphafeh}
\end{figure}
Based on the photospheric chemical composition and Milky Way kinematics of KIC\,10001167, \citet{montalban2021} classified it as a member of the Milky Way's in-situ high-[$\alpha$/Fe] disk.
In Fig.~\ref{fig:article_alphafeh}, KIC\,10001167 is shown in the broader context of Milky Way giants observed by APOGEE DR17, where its combination of [Fe/H] and high $\rm [\alpha/Fe]$-levels makes it clearly distinct from stars accreted from other galaxies \citep{Helmi2020}, and therefore a prototypical in-situ star. We note that its location in the [Mg/Mn]-[Al/Fe] plane, which has been shown to clearly separate in-situ disk stars from those born ex-situ \citep{Das2020}, further demonstrates its membership to the in-situ thick disc.

To provide additional and independent chemical constraints, we measure the iron abundance and detailed abundances of nine other elements using high-resolution spectroscopic data from the separated FIES spectrum of the RG. 
We find a logarithmic iron abundance of $\rm [Fe/H] = -0.73\pm 0.10$, and an alpha-element enhancement of $\rm [\alpha/Fe]=+0.37 \pm 0.10$. This is compatible with APOGEE DR17, and an independent confirmation of the system's chemical association to the old in-situ disk.
Further details on the methodology are available in Section~\ref{sec:specanal}.

Moreover, using  astrometric constraints from \emph{Gaia} DR3 \citep{gaia2022}, with our independent systemic RV measurement obtained from the FIES spectra, we find that the star's Galactic orbit, particularly its eccentricity of ${0.42} \pm 0.02$ and orbital circularity of $\rm L_z/L_c = 0.8$, with $L_z$ being the orbital azimuthal angular momentum, and $L_c$ being the equivalent value for a circular orbit with the same energy, is compatible with an origin in the old in-situ population \citep[e.g.][and references therein]{Chandra2023}.

Using photometric constraints from \emph{Gaia} DR3 and 2MASS \citep{2MASS2006}, we measure a largely model-independent effective temperature and angular diameter of the RG with the infra-red flux method (IRFM).
The effective temperature of $\rm 4625\pm 29_{stat} \pm 30_{syst}\, K$ is compatible with both our spectroscopic analysis and APOGEE DR17.

By combining our photometric IRFM measurement of angular diameter with the \emph{Gaia} DR3 parallax, we measure a radius of $12.82\pm 0.30_{\rm stat}\pm 0.24_{\rm syst}$ $\rm R_{\odot}$ for the RG, where the systematic uncertainty includes the potential effect of the binary orbit on the parallax.

Our photospheric constraints for the RG can be found in Table~\ref{table:RGdata}, while a table of detailed abundances can be found in Appendix Table~\ref{table:abundances}.

\subsection{Eclipsing binary analysis}\label{article:eb}
\begin{figure}
    \centering
    \includegraphics[width=\columnwidth]{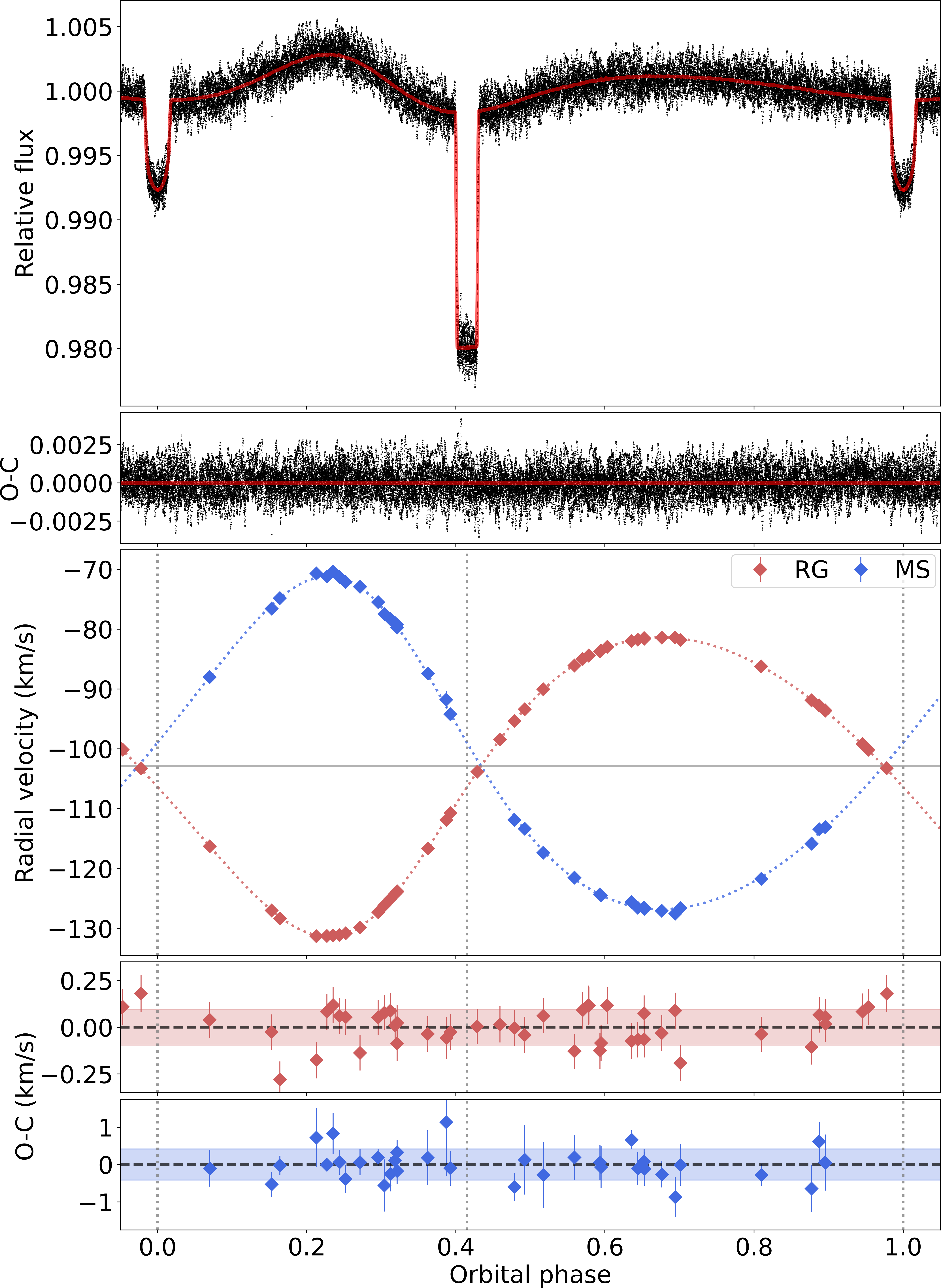}
    \caption{Top: Binary signal in the light curve, along with the best-fit PHOEBE 2 model. Bottom: The RV measurements and dynamical RV curves for the two stellar components. Dotted vertical lines indicate the location of eclipses. The Observed-Calculated (O-C) panels refer to the model-subtracted residuals.}
    \label{fig:article_eb}
\end{figure}

The \emph{Kepler} light curve shows clear eclipses of two stellar components, and a signal from tidal deformation and Doppler beaming of the RG.
This is demonstrated in Fig.~\ref{fig:article_eb} together with our spectroscopic follow-up RVs from FIES.

To measure orbital and stellar parameters, we perform combined eclipsing binary light curve and RV analysis with two independent programs, JKTEBOP \citep{Southworth2013} for analysis modelling only the light curve eclipses, and PHOEBE 2 \citep{conroy2020} for analysis that incorporate tidal deformation and Doppler beaming.
While the two codes have different underlying assumptions, chiefly on stellar sphericity, we find an agreement on the RG radius of $0.4\%$ ($0.4\sigma$) between them. Further details on the two analyses can be found in Section~\ref{sec:binary}.
All the measured orbital and stellar parameters are found in Appendix Table~\ref{table:EBdata}.

With the eclipsing binary analysis, we measure the RG mass as $0.9337\pm 0.0077 \,\mathrm{M_{\odot}}$ ($0.8\%$), and its radius as $13.03\pm 0.12 \,\mathrm{R_{\odot}}$ ($0.9\%$). 
In Sect.~\ref{sec:jktebop}, we furthermore estimate a potential systematic uncertainty of $0.7\%$ for the RG radius, from assumptions related to our treatment of the stellar atmosphere and limb darkening, and its value is quoted in Table~\ref{table:RGdata}.

\begin{figure}
    \centering
    \includegraphics[width=\columnwidth]{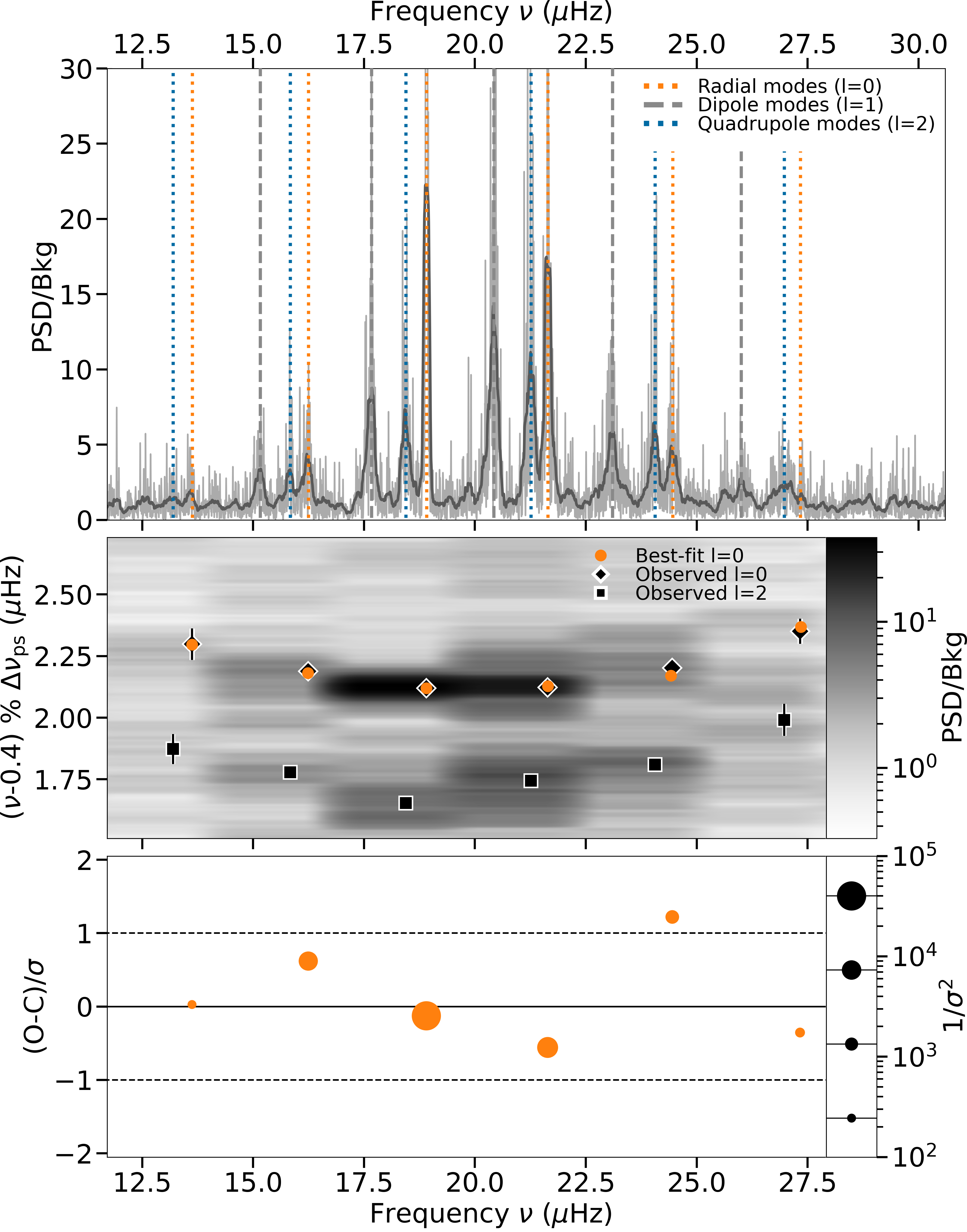}
    \caption{Top: Frequency-power spectrum divided by the granulation background, both in original (light) and uniformly smoothed (dark, window=$0.15\mu$Hz). Vertical lines highlight the observed radial ($\ell$=0), dipole ($\ell$=1) and quadrupole ($\ell$=2) modes. Middle: Échelle diagram, with the axes flipped for illustration, showing observed radial $\ell=0$ and quadrupole $\ell=2$ frequencies, and best-fit frequencies from our reference radial mode fit. Heatmap data is uniformly smoothed with window=$0.075\mu$Hz. Bottom: Statistical significance of the Observed-Calculated (O-C) residuals relative to measurement uncertainty $\sigma$. Marker-size has been re-scaled (in $\log_{10}$) to demonstrate the statistical weight $1/\sigma^2$ applied to each observed frequency in the asteroseismic inference.}
    \label{fig:article_psd}
\end{figure}

\begin{figure}
    \centering
    \includegraphics[width=\columnwidth]{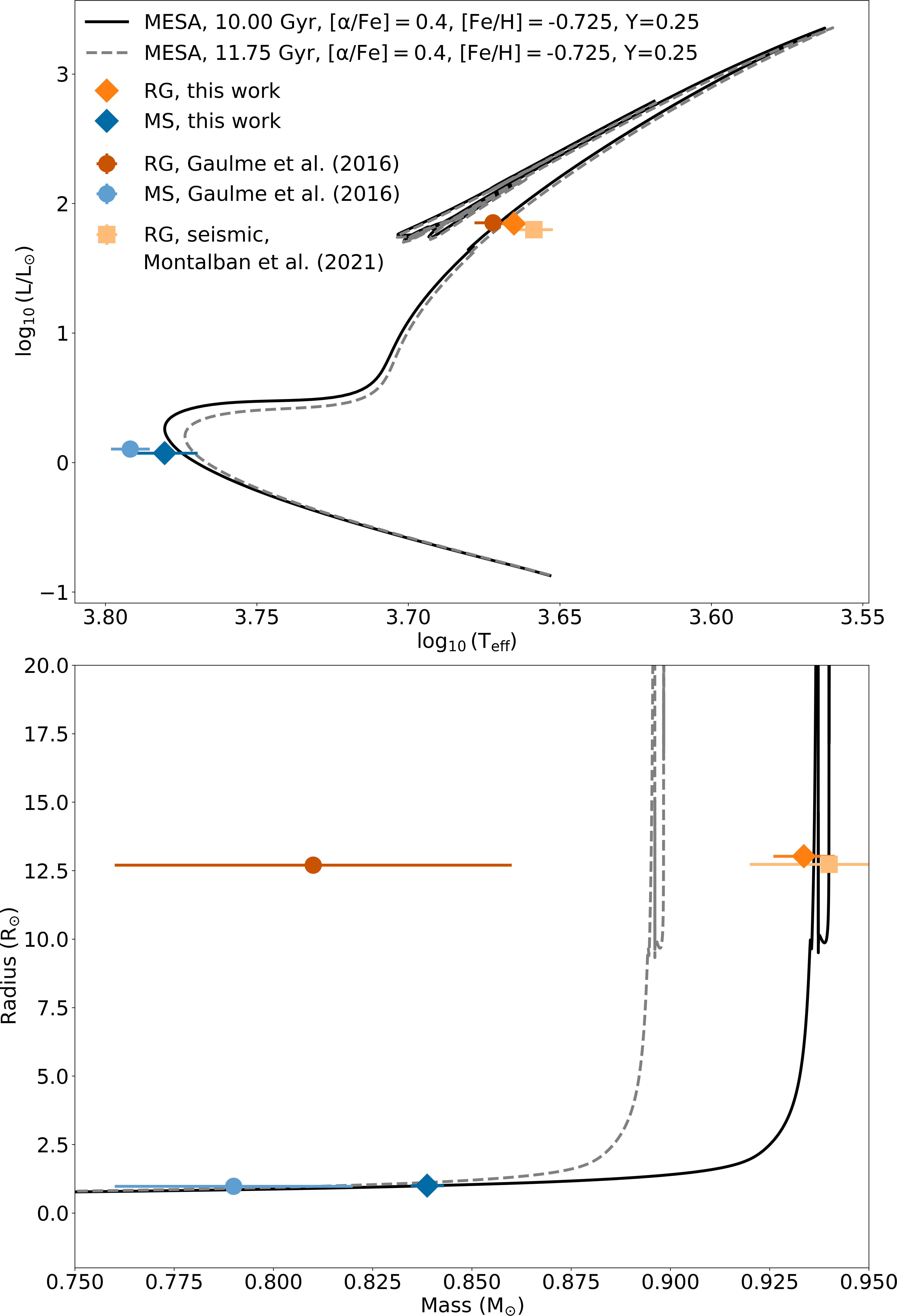}
    \caption{Top: Hertzsprung-Russell diagram with luminosity of RG and MS from eclipsing binary radius and IRFM temperature (this work). Included is two isochrones calculated from the MESA grid of stellar models used in this paper. Also shown are eclipsing binary measurements from \citet{Gaulme2016}, and asteroseismic inference of the RG from \citet{montalban2021}. Bottom: The mass and radius of the same sources, along with the same isochrones. All markers have error-bars in x and y, but in some cases they are smaller than the marker.}
    \label{fig:article_hrmr}
\end{figure}

\begin{figure}
    \centering
    \includegraphics[width=\columnwidth]{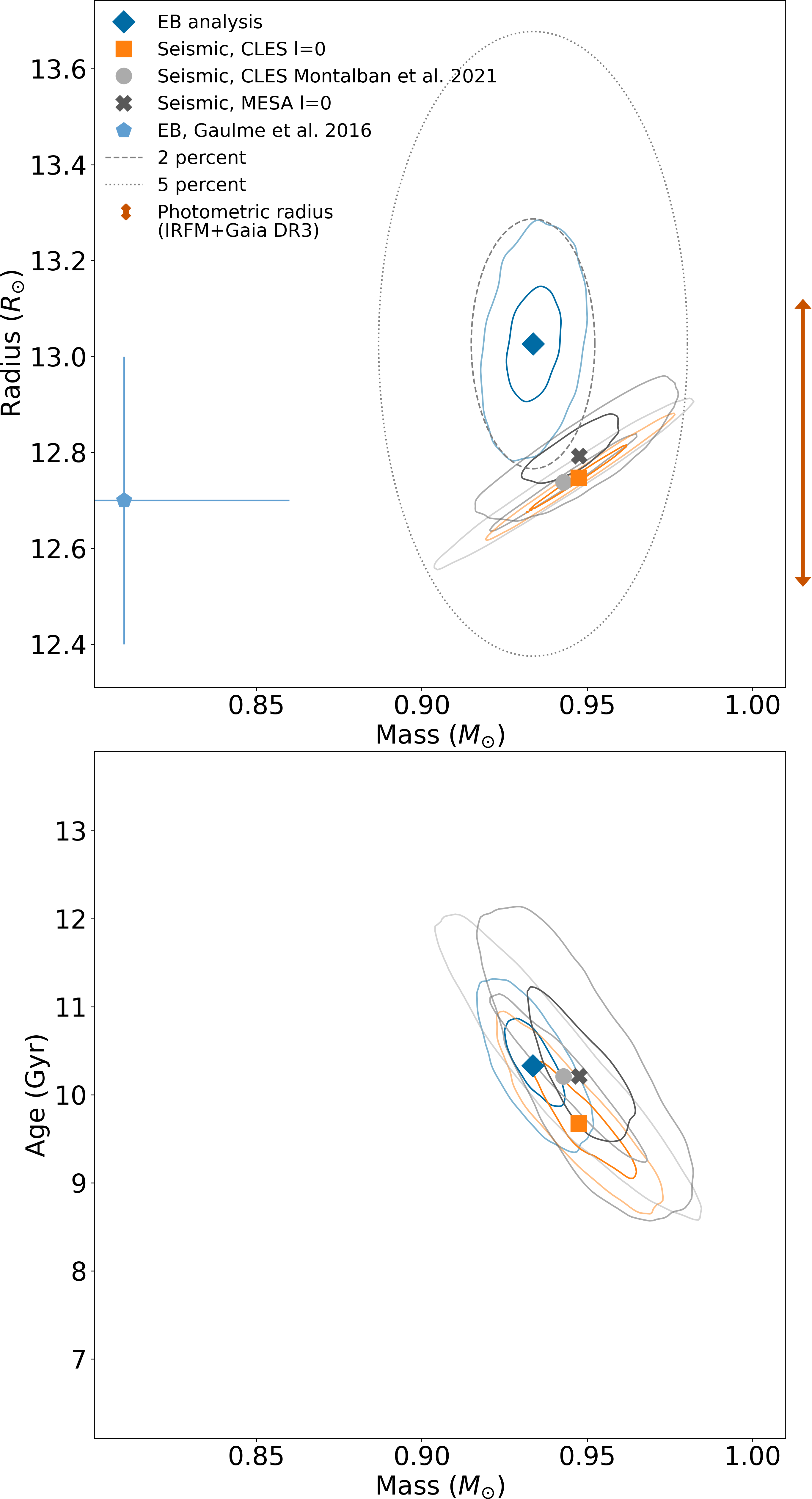}
    \caption{Top: RG Mass vs. Radius measurements from this work and the literature. Contours represent one and two sigma. As guides, circles have been drawn around our eclipsing binary measurement representing $2$ and $5$\% radial difference in mass and radius. The arrow represents $\pm 1\sigma$ for the radius measurement with the infra-red flux method and \emph{Gaia} DR3 parallax.  Bottom: Mass vs. Age from stellar model inferences.}
    \label{fig:article_mrfit}
\end{figure}

\subsection{Asteroseismic constraints and modelling}\label{article:seis}
Figure~\ref{fig:article_psd} shows the frequency--power spectrum of the pre-processed lightcurve (see Section~\ref{sec:asteroseismology}).
The light curve from KIC10001167 presents a rich spectrum of overtones of solar-like oscillations from the RG, modes that are stochastically excited and intrinsically damped by near-surface convection. The modes may be decomposed onto spherical harmonics of angular degree $\ell$. Overtones of radial ($\ell = 0$), dipole ($\ell = 1$) and quadrupole ($\ell = 2$) modes are clearly seen.  The structure of dipolar modes is informative of the evolutionary state of the star \citep[e.g.][]{Bedding11, Mosser2014}, supporting previous analyses \citep{Elsworth2019, Pinsonneault2018}  which demonstrated that this star is in the red giant branch phase, i.e. in the hydrogen-shell burning phase which follows the exhaustion of hydrogen in the stellar core (further details can be found in Appendix~\ref{sec:evol_state}).
We measure the frequencies of individual radial and quadrupole modes using well established data analysis procedures \citep{corsaro2020, nielsen2021, Arentoft17}, see Sect.~\ref{sec:asteroseismology} for further details.

We then infer stellar properties using individual-mode frequencies and photospheric parameters from the IRFM and our optical spectroscopy, or APOGEE DR17, as observational inputs in AIMS \citep{Rendle2019}. Here we extend the work presented in \citet{montalban2021} by exploring the impact of using different combinations of observational constraints and uncertainties in the modelling.
We find a radius of $12.748\pm0.068_{\rm stat}\pm0.055_{\rm syst}$ $\mathrm{R_\odot}$, a mass of $0.947\pm0.015_{\rm stat}\pm0.009_{\rm syst}$ $\mathrm{M_\odot}$, and an age of $9.68\pm0.64_{\rm stat}\pm0.56_{\rm syst}$ Gyr. This is consistent with the asteroseismic inference of \citet{montalban2021}, where they adopted a reduced set of oscillation frequencies and spectroscopic constraints from an earlier APOGEE data release.

The asteroseismic results are found to be robust against the systematic effects explored in Appendix~\ref{sec:inference-details}; using different stellar structure and pulsation codes, different model temperature scales, including Gaia-based luminosity, adopting abundances from APOGEE or from FIES, and including quadrupole modes as observational constraints.
We also perform an inversion for the mean stellar density following the approach described in \citet{buldgen2019} finding results consistent with those from the forward modelling approach (see Appendix~\ref{sec:inversion}).  

In Appendix~\ref{sec:tides}, we argue that it is unlikely that tidal effects could have caused a significant bias in the asteroseismically determined radius.

In Appendix~\ref{sec:mass-loss}, we explore potential systematic effects related to mass loss during the RGB and find them to be similar to the currently adopted systematic uncertainty on age.

When using the mass and radius measurements from the eclipsing binary analysis as observational constraints instead of the seismic parameters, the recovered age is $10.33\pm 0.48_{\rm stat}\pm 0.38_{\rm syst}$ Gyr, $1\sigma$ ($6\%$) consistent with the asteroseismic inference.

\begin{table}
\caption{\label{table:RGdata} Measurements of the red giant in KIC\,10001167.}
\centering
\begin{tabular}{lll}
\hline\hline
Quantity & \multicolumn{2}{c}{Value} \\
\hline
$\rm T_{eff, IRFM}$ (K)                                      &\multicolumn{2}{c}{$4625 \pm 29_{\rm stat} \pm 30_{\rm syst}$}\\
$\rm [Fe/H]_{FIES}$ (dex)                                          &\multicolumn{2}{c}{$-0.73 \pm 0.10$}\\
$\rm [\alpha/Fe]_{FIES}$ (dex)                                 &\multicolumn{2}{c}{$0.37 \pm 0.10$}\\
\hline
\hline
Mass$_{\rm dyn} ~(\rm M_{\odot})$                        &\multicolumn{2}{c}{0.9337 $\pm$ 0.0077}\\
Mass$_{\rm sis,l=0} ~(\rm M_{\odot})$                        &\multicolumn{2}{c}{$0.947 \pm 0.015_{\rm stat} \pm 0.009_{\rm syst}$}\\
\hline
Radius$_{\rm dyn} ~(\rm R_{\odot})$                  &\multicolumn{2}{c}{$13.03 \pm 0.12_{\rm stat} \pm 0.09_{\rm syst}$}\\
Radius$_{\rm sis,l=0} ~(\rm R_{\odot})$                  &\multicolumn{2}{c}{$12.748 \pm 0.068_{\rm stat} \pm 0.055_{\rm syst}$}\\
Radius$_{\rm IRFM}~(\rm R_{\odot})$                 &\multicolumn{2}{c}{$12.82 \pm 0.30_{\rm stat} \pm 0.26_{\rm syst}$}\\
\hline
$\rm age_{\rm dyn} ~(Gyr)$                       &\multicolumn{2}{c}{$10.33\pm0.48_{\rm stat} \pm0.38_{\rm syst}$}\\
$\rm age_{sis, l=0} ~(Gyr)$                      &\multicolumn{2}{c}{$9.68 \pm 0.64_{\rm stat} \pm 0.56_{\rm syst}$}\\
\hline
\end{tabular}
\end{table}

\section{Discussion and conclusions}\label{discussion}
KIC\,10001167 represents an exceptionally well-constrained binary system, prototypical of stars formed in the Milky Way's in situ high-$\rm [\alpha/Fe]$ disk at an iron abundance $\rm [Fe/H] \simeq -0.7$ and $\rm [\alpha/Fe] \sim 0.3-0.4$, hence predating significant enrichment by Type Ia supernovae \citep[e.g., see][]{Matteucci1986}.

Thanks to our radial velocity monitoring program, we have collected data that enable the mass of the red giant star in KIC\,10001167 to be measured with a precision of 1\%. Since the star is a low-luminosity giant, model-independent knowledge of its mass enables an age inference from stellar models of $10.3 \pm 0.5_{\rm stat} \pm 0.3_{\rm syst}$ Gyr, independent of the star's asteroseismic constraints.

In Fig.~\ref{fig:article_hrmr}, we illustrate with two representative isochrones how the evolved nature of the RG ensures that knowledge of its mass is directly informative of the age of the system, regardless of the star's luminosity, temperature and radius.


Fig.~\ref{fig:article_mrfit} illustrates the comparison between mass, radius and age for our measurements and the literature. We find that our asteroseismic mass measurements based on detailed seismic modelling agree with the dynamical mass to a level of $1.4\%$, which corresponds to $0.8\sigma$, or $0.3\sigma$ when accounting for systematic sources of uncertainty. Furthermore, Fig.~\ref{fig:article_mrfit} demonstrates that this $\sim 1\sigma$ difference in measured mass matches directly as expected to the $\sim 1\sigma$ difference in inferred age between the two independent sets of observational constraints (eclipsing binary and asteroseismic).

The asteroseismically inferred photospheric radius is found to agree within $2.1\%$ with the dynamical radius ($2.1\sigma$). However, if we consider the systematics in both analyses, the difference between the measurements may be as low as $1.1\sigma$. 
Further research is needed to ascertain the significance of this difference, which also affects the inferred mean density. This includes a thorough evaluation of the systematic errors associated with current treatments of the stellar atmosphere in eclipsing binary models for giants, especially the limb darkening prescriptions.
The independent photospheric radius obtained with the IRFM and the \emph{Gaia} DR3 parallax, with its precision of $2.3\%$, is compatible with both measures within $1\sigma$.

While independent checks of asteroseismically inferred radii (and hence distances) can also be performed using precise \emph{Gaia} parallaxes for thousands of stars \citep[e.g., see][and references therein]{Khan2023}, high accuracy comparisons between independent stellar mass determinations are unique to binaries. In this context, the percent-level agreement on mass we obtain for KIC\,10001167 demonstrates that asteroseismic inferences using individual oscillation modes provide a method for achieving not only high-precision \citep[2\%, see][]{montalban2021}, but also high-accuracy measurements of masses and ages for thousands of the oldest RGB stars in the Milky Way. We have thus demonstrated that asteroseismology offers the opportunity to accurately probe the assembly history of the Milky Way at early cosmic times. Asteroseismology could further be used to establish a fundamental training set for data-driven techniques, enabling the inference of ages for millions of stars with a truly improved temporal resolution that we are currently lacking.

Extending the sample of fundamental mass and age calibrators remains critically important. Currently, KIC\,10001167 is the only old disk red giant hosting binary with the necessary asteroseismic and dynamical constraints for such detailed comparisons. However, the upcoming mission PLATO \citep{Rauer2014,Rauer2024, Miglio2017} has the potential to increase this sample, to which Gaia will also contribute with astrometric binaries \citep{Beck2024}.
Finally, further tests of the reliability of stellar models, as well as the asteroseismic age and mass scales in different Galactic environments and stellar clusters—as envisioned by the HAYDN mission \citep{Miglio2021b}—will help further enhance the time resolution with which we can study the history of the Milky Way.


\begin{acknowledgements}\\

JST acknowledge support from Bologna University, "MUR FARE Grant Duets CUP J33C21000410001".

KB, JM, MT, GC, AM, MM acknowledge support from the ERC Consolidator Grant funding scheme (project ASTEROCHRONOMETRY, \url{https://www.asterochronometry.eu}, G.A. n. 772293).

Funding for the Stellar Astrophysics Centre is provided by The Danish National Research Foundation (Grant agreement no.: DNRF106).

Based on observations made with the Nordic Optical Telescope, owned in collaboration by the University of Turku and Aarhus University, and operated jointly by Aarhus University, the University of Turku and the University of Oslo, representing Denmark, Finland and Norway, the University of Iceland and Stockholm University at the Observatorio del Roque de los Muchachos, La Palma, Spain, of the Instituto de Astrofisica de Canarias.

This research is also supported by work funded from the European Research Council (ERC), the European Union’s Horizon 2020 research, and innovation programme (grant agreement n◦803193/BEBOP).

DJ acknowledges support from the Agencia Estatal de Investigaci\'on del Ministerio de Ciencia, Innovaci\'on y Universidades (MCIU/AEI) and the European Regional Development Fund (ERDF) with reference PID-2022-136653NA-I00 (DOI:10.13039/501100011033). DJ also acknowledges support from the Agencia Estatal de Investigaci\'on del Ministerio de Ciencia, Innovaci\'on y Universidades (MCIU/AEI) and the the European Union NextGenerationEU/PRTR with reference CNS2023-143910 (DOI:10.13039/501100011033).

GB acknowledges fundings from the Fonds National de la Recherche Scientifique (FNRS) as a postdoctoral researcher.

This work was supported by the UK Science and Technology Facilities Council under grant number ST/Y002563/1.
This paper includes data collected by the Kepler mission. Funding for the Kepler mission is provided by the NASA Science Mission directorate.
\end{acknowledgements}

%
%

\bibliographystyle{aa}
\bibliography{thomsen}

\begin{appendix} 
\section{Radial velocity analysis details}\label{sec:rv_details}
Detailed outputs from the separation and RV analysis are reported in Appendix Table~\ref{table:rvs}.
For spectral separation, each spectrum is weighted according to its exposure time, as well as component RV separation and eclipse occurrence. 
We calculate RV corrections based on the wavelength positions of telluric lines, by cross-correlating the strong telluric lines at 6865--6925Å.
We estimate RV uncertainties by combining in quadrature the following uncertainty estimates: 1.
Internal uncertainty from stellar RV measured on smaller wavelength intervals, corrected for a linear instrumental trend with wavelength for the RG likely caused by systematics in either the instrumental line profile or the wavelength solution (RG mean $21$ m/s, MS mean $0.39$ km/s).
2. Local night-to-night scatter in the cross-correlation of the ThAr spectra, not considering long-term trends ($6-9$ m/s).
3. Cross-validation uncertainty estimate for the telluric corrections ($11-44$ m/s).
4. A fixed RV jitter uncertainty from correlated noise, evaluated from the best-fit JKTEBOP RG residuals ($91$ m/s).
Excluding (including) jitter, we find a mean RV uncertainty of $29$ ($96$) m/s for the RG and $0.48$ ($0.49$) km/s for the MS star.
\section{Radial velocity verification}\label{sect:rv-residual}
In Sect.~\ref{sec:binary}, we find the presence of an additional signal in the RVs of the RG beside classical two-body Keplerian motion. We rule out long-term variations in the Keplerian orbit as the sole cause of this (e.g. eccentricity change, period change, or tidal apsidal motion) by trial fitting with perturbed models. 
Line profile variations for the RG could potentially explain some of the behavior. Non-Keplerian Doppler shifts e.g. pulsations, or a perturbing long-period circumbinary component, can each account for parts of the signal but cannot be proven without a longer baseline.

We compare our measured RVs to a newly developed method, which allows us to measure both components of high-contrast binaries using cross-correlation with a line mask \citep{Sebastian24b}. In this method we analyse the FIES data in two steps. In a first step, we make use of the high contrast ratio and measure the radial velocity of the RG component alone using a cross correlation function (CCF) with a K6 line mask\footnote{We use ESPRESSO\citep{pepe21} line masks, which have been published on \hyperlink{https://www.eso.org/sci/software/pipelines/espresso/espresso-pipe-recipes.html}{https://www.eso.org}}. We then measure the radial velocity (RV), CCF contrast, full-with-half-maximum (FWHM) and bisector span \citep{queloz01} by fitting a Gaussian to each CCF. We then use {\texttt kima} \citep{Faria18,Baycroft23}, which applies a Keplerian fit using a diffusive nested sampler to the measured RVs. Here we apply a Gaussian prior for the system period, measured from photometric data, a log-uniform prior for the semi-amplitude ($K_1$) between 24 and 27\,${\rm km s^{-1}}$, as well as wide priors for the eccentricity (ecc) and argument of periastron ($\omega$).

The measured radial velocity values for the RG are $1\sigma$ compatible with our results using the broadening function (BF) technique in the main analysis. To avoid the bluest and most noisiest orders, we analyse the spectra between 461.1\,nm and 644.6\,nm. We also verify that the measured value for $K_{\rm RG}$ is consistent for different orders in this range. The orbital parameters, of the RG, obtained from CCF fitting are displayed in Table~\ref{tab:CCF_res}. The RV residuals from the simple Keplerian fit do indicate a possible trend, as well as significant short-term variation for newer spectra, also seen in our BF analysis. However, it is not clear with the current baseline if the trend is due to short-term variations affecting the system velocity measurement, or true long-term variation. This trend could indicate a physical companion, or be caused by either activity or pulsations of the giant. We do not find significant bisector variation that could attribute the residuals to activity or a luminous companion (a luminous companion was also independently investigated in Appendix~\ref{sec:spec-tlight}). We were unable to obtain a conclusive indication of apsidal motion, period/eccentricity variation, or mass loss. A longer base line would be necessary to securely conclude on this trend. To account for the RV residuals, we add quadratically a base-systematic error of 90\,${\rm m s^{-1}}$ to the fit uncertainties from the Gaussian fit, equivalent to the main analysis. Before analysing the MS star, we remove two spectra, which are more than $10\sigma$ outliers in FWHM (BJD: 2459047.5, \& 2460032.7).

\begin{figure}[ht]
\centering
\includegraphics[width=0.42\columnwidth]{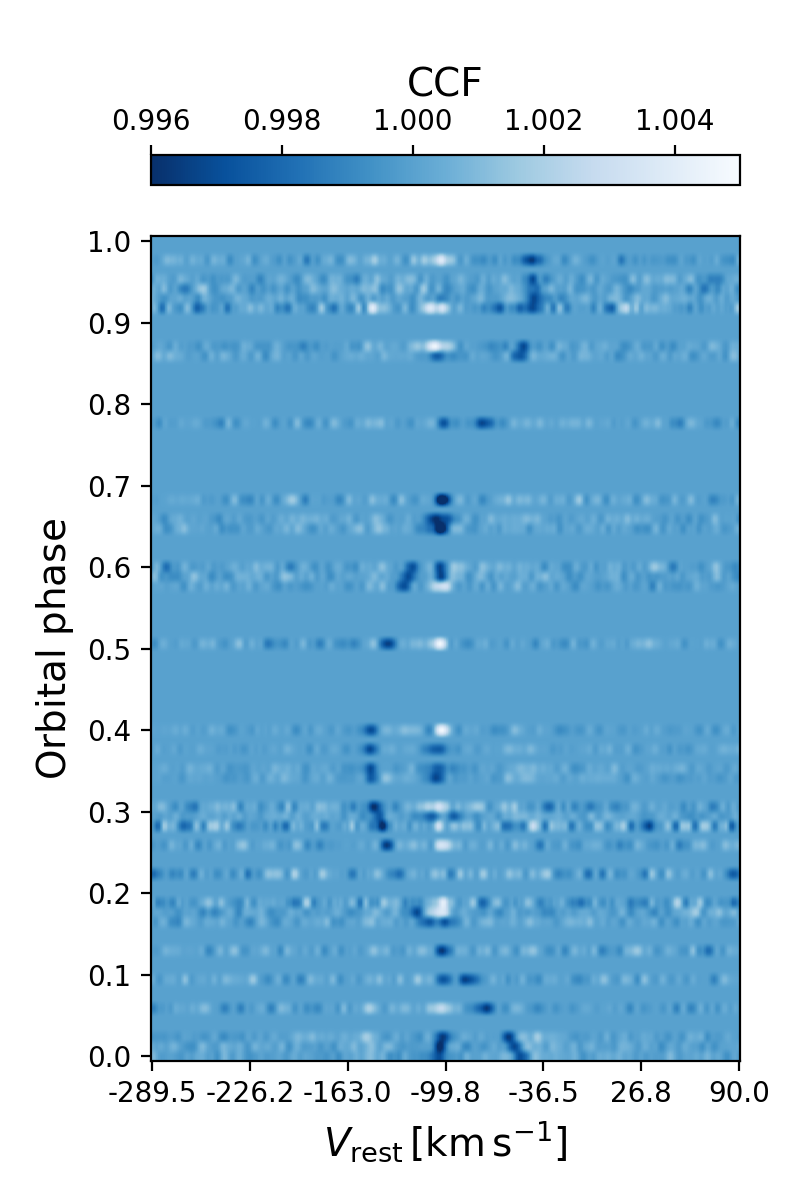}
\includegraphics[width=0.55\columnwidth]{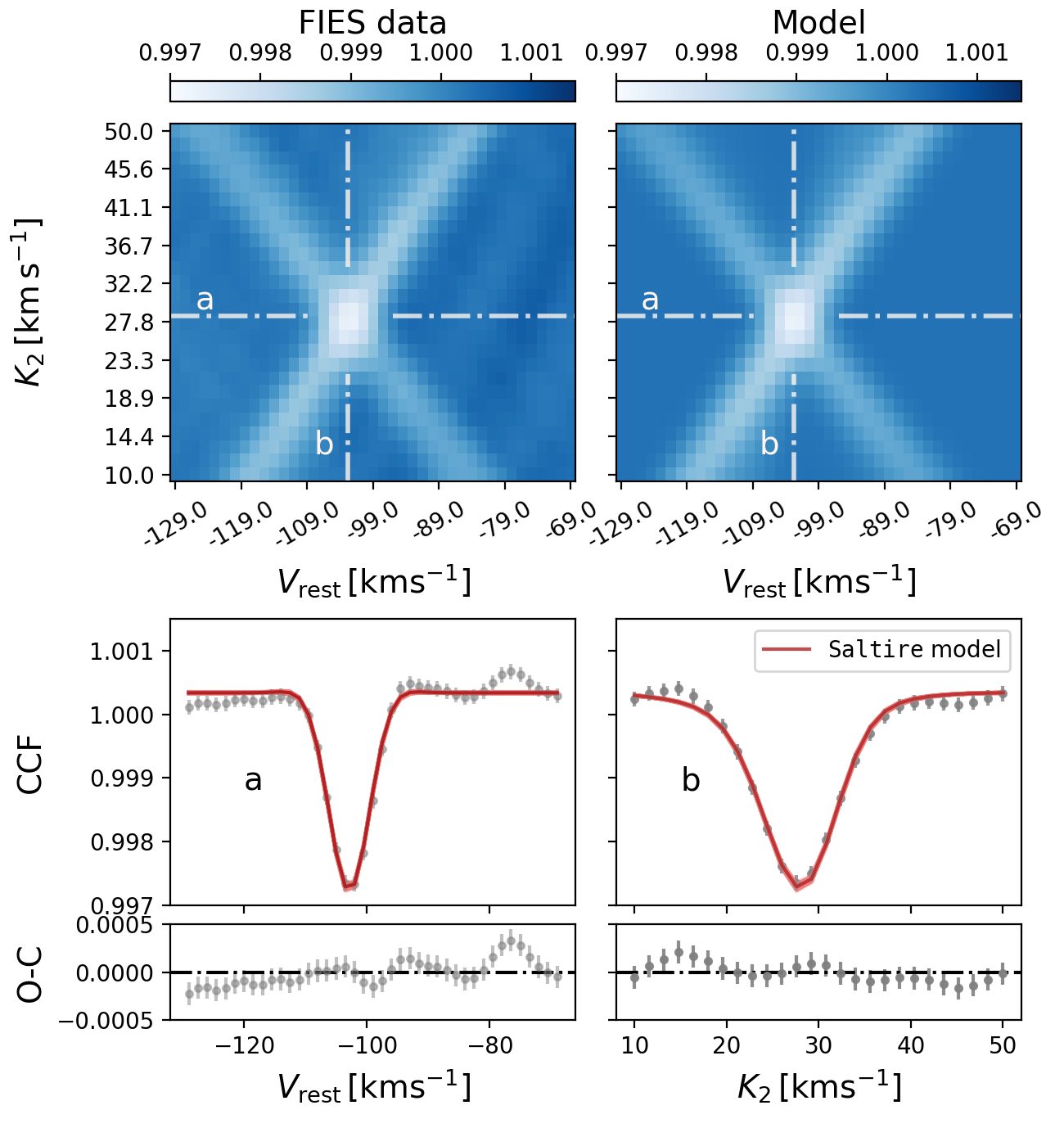}
\caption{Cross-correlation functions of SVD detrended spectra. Left panel: As a function of the orbital phase (0 is the time of periastron) in the RG rest frame, with the MS companion clearly detected. Right upper panels: CCF maps fom the K-focusing method. A 26.6-$\sigma$ CCF signal marks the MS star's orbit being aligned in the MS's rest-frame. The best fitting {\textit Saltite} model is used to measure the MS star semi-amplitude and rest-velocity. White lines (a \& b) mark sections shown in the Right lower panels: Gray data points are CCF data, error bars represent the overall jitter from the two-dimensional fit.}
\label{fig:CCF_sec}
\end{figure}

\begin{table}
\centering
\caption{\label{tab:CCF_res}Binary parameters obtained from CCF analysis.}
\begin{tabular}{ll}
\hline
\hline
RG parameters & Value \\
\hline
$K_{\rm RG}$ [${\rm km s^{-1}}$] & 24.985$\pm$0.022 \\
$V_{\rm rest,RG}$ [${\rm km s^{-1}}$] & -102.525$\pm$0.075 \\
$T_{0,peri}$ [BJD] & 2459156.57$\pm$0.22 \\
$P$ [d] & 120.38994$\pm$0.00059 \\
ecc & 0.15773$\pm$0.00095 \\
$\omega$ & 3.6910$\pm$0.0058 \\
CCF contrast [$\%$] & 21.04$\pm$0.03 \\
${\rm FWHM_{\rm RG}}$ [${\rm km s^{-1}}$] & 9.398$\pm$0.019 \\
\hline
MS parameters & Value \\
\hline
$K_{\rm MS}$ [${\rm km s^{-1}}$] & 28.00$\pm$0.12 \\
$V_{\rm rest,MS}$ [${\rm km s^{-1}}$]& -102.806$\pm$0.067 \\
CCF contrast [$\%$] & 0.317$\pm$0.036 \\
${\rm FWHM_{\rm MS}}$ [${\rm km s^{-1}}$]& 7.74$\pm$0.31 \\
\hline
\end{tabular}
\end{table}

In a second step, we use the RV measurements to align all remaining 43 FIES spectra into the rest frame of the RG star. We then detrend \citep{Sebastian24b} the spectra by removing on average the first three components of a singular value decomposition (SVD, \citet{kalman1996}). This should effectively remove the lines of the RG star. The detrended spectra are then cross-correlated with a G9 line mask. Figure~\ref{fig:CCF_sec} shows the CCF in the primaries rest frame. The trail of the MS star is clearly visible, as well as some residuals from the RG. These residuals are most likely caused by variations of the RG's absorption spectrum during the $\sim4.5\,$yr of observation. Systematic residuals from the RG were also found in the spectral separation performed in the main analysis, supporting the non-artefact origin. For the analysis of the companion, we therefore exclude all spectra where the radial velocity difference of both binary components is less than 20\,$\rm km s^{-1}$. Since the companion signal is clearly visible in the CCF, we also identify three spectra which show very noisy CCF's, and therefore exclude them from the analysis.      
For the 32 selected and detrended spectra, we use the K-focusing method to measure the MS companion semi-amplitude \citep{Sebastian24a}. In this process, we keep the orbital parameters of the RG fixed and sample the MS semi-amplitude ($K_2$) in steps of 1.5\,$\rm km s^{-1}$ from 10 to 50\,$\rm km s^{-1}$ and the companion rest velocity ($V_{\rm rest,2}$) in steps of 1.5\,$\rm km s^{-1}$ from -130 to -70\,$\rm km s^{-1}$. Figure~\ref{fig:CCF_sec} shows the cross-correlation map of the companion, showing a 26.6-$\sigma$ detection. We use the {\texttt Saltire}\footnote{Saltire python code and documentation is available on \href{https://github.com/dsagred/Saltire}{Github}. } model to fit the map, which allows us to obtain precise parameter measurements of the MS orbit. The model assumes that the CCF of the MS companion follows a Gaussian shape, thus we can measure $K_2$ and $V_{\rm rest,2}$, but also FWHM and relative CCF contrast of the companion's mean line profile. To fit the data, we use a Markov Chain Monte Carlo (MCMC) implementation in {\texttt Saltire} to sample the posterior distribution for each of the fit parameters with 21,000 samples. The sections through the map in Figure~\ref{fig:CCF_sec} show some deviations from this shape. The individual spectrum CCFs are also more noisy than the BF. This is likely due to the same as the K-focusing map sections: Additional statistical noise (the line mask is optimized for m/s exoplanet detection rather than SNR), the CCF has some minor peak-pulling behavior and side-lobes compared with the BF, and the left-over spectrum has summed-up residuals from the SVD \citep{Sebastian24a} which are different from the non-SVD separation approach used in the main analysis.
To measure the systematic uncertainties from this analysis, we split the data into 4 partial samples with four spectra each. We then repeat the analysis for each of them and adopt the RMS error as systematic uncertainty\citep{Sebastian24b}, which we add to the fit uncertainties from the {\texttt Saltire} fit. The final results and uncertainties are reported in Table~\ref{tab:CCF_res}.

The most important difference between the main RV analysis and this is on the semi-amplitude of the MS star, at $1.6\sigma$. The SNR of the MS star in the spectra is very low, near $\sim1$ for the CCF analysis, and the detections are primarily noise-dominated, with some additional residuals from the SVD and CCF. As the BF analysis has a higher SNR detection for the MS star in all spectra, the difference of less than $2\sigma$ between two different methods, with completely different weighing of the data, can be reasonably argued as statistical. When performing a similar two-step analysis with the broadening function method and spectral separation following \citet{Gonzalez2006}, by first fitting the RG individual RVs, and then fitting the MS orbital semi-amplitude with all other orbital elements fixed, we obtain $K_{\rm MS} = 27.87$ km/s, $0.5\sigma$ different from the main analysis and $1\sigma$ from the CCF analysis.
\section{Spectroscopy, choice of line-list}\label{sect:linelist}
The line-list used by the Gaia-ESO survey\citep{stetson08} has a large number of blends and irregularly shaped lines for the RG. This is the reason we used the manually selected line-list given in \citet{Slumstrup2019}. The line-list oscillator strengths of each absorption line has been calibrated on a solar spectrum obtained with the same spectrograph.
\begin{table*}
\centering
\caption{\label{table:abundances}
Spectroscopic atmospheric parameters, and abundances, for the RG in KIC\,10001167, both ours (FIES) and from the APOGEE DR17 near-infrared survey. See text for details on the analysis. Solar reference for the FIES abundances is \citet{Asplund2009}, while for APOGEE DR17 it is \citet{grevesse2007}. The number of lines used for each FIES abundance measurement is referred to as "n" (including nFeI(I)). $\xi$ refers to the atmospheric micro-turbulent velocity. The abundance uncertainty estimates for FIES are the standard deviation of the individual lines against the mean, and not the standard errors.
}
\begin{tabular}{lccc}
\hline
\hline
Quantity  & APOGEE DR17\tablefootmark{a}  & FIES, astrophysical $\log(\rm gf)$ & FIES, laboratory $\log(\rm gf)$ \\
& & \citep{Slumstrup2019} & \citep{stetson08} \\
\hline
$T_{\rm eff} (\rm K)$ & $4598 (8)$   &   $4622 (36)$ &  $4698 (68)$\\
$\log g \rm (cgs)$      & $2.209$   & fixed $2.20$   & fixed $2.20$ \\
$\rm [FeI/H] (dex)$  & $-0.677 (9)$\tablefootmark{b} &   $-0.726 (66)$ &  $-0.672 (99)$\\
$\rm [FeII/H] (dex)$ &  &   $-0.786 (67)$ &  $-0.877 (74)$\\
$\rm [\alpha/Fe] (dex)$ & $0.277 (8)$ &   $0.369 (69)$ & $0.319 (100)$\\
$\xi (\rm km s^{-1})$ & &     $1.230 (60)$ &  $1.290 (70)$\\
nFeI & & 73 & 68 \\
nFeII & & 12 & 12 \\
\hline
$\rm [NaI/Fe]$ &$0.067 (88)$&   $0.267 (67)$, n=2       & $0.230 (99)$, n=2 \\
$\rm [MgI/Fe]$ &$0.339 (15)$&   $0.412 (68)$, n=3       & $0.463 (99)$, n=2 \\
$\rm [AlI/Fe]$ &$0.302 (21)$&   $0.404 (78)$, n=2 &  $0.419 (102)$, n=2\\
$\rm [SiI/Fe]$ &$0.248 (16)$&   $0.379 (66)$, n=5       & $0.190 (103)$, n=7 \\
$\rm [CaI/Fe]$ &$0.194 (19)$& $0.254 (84)$, n=5   & $0.249 (100)$, n=3 \\
$\rm [TiI/Fe]$ &$0.150 (23)$&   $0.429 (95)$, n=11      & $0.373 (108)$, n=8 \\
$\rm [TiII/Fe]$&$0.375 (83)$&   $0.379 (76)$, n=2       & $0.305 (99)$, n=2 \\
$\rm [CrI/Fe]$ &$-0.055 (54)$&   $0.106 (80)$, n=5       & $0.036 (117)$, n=5 \\
$\rm [Mn/Fe]$   &$-0.150 (20)$ & &\\
$\rm [NiI/Fe]$ &$0.087 (16)$&   $0.102 (86)$, n=8       & $0.081 (121)$, n=9 \\
\hline
\end{tabular}
\tablefoot{
\tablefoottext{a}{For APOGEE DR17, all measurements are "overall" abundances $\rm X$ instead of uniquely atomic neutral $\rm XI$ or single-ionized $\rm XII$, with the exception of $\rm [TiII/Fe]$ (atomic single-ionized). For the rest of the elements, whether the APOGEE abundances are marked $\rm [X/Fe]$ or $\rm [XI/Fe]$ has no bearing on the actual state.}
\tablefoottext{b}{This is the quoted [M/H] value of the overall APOGEE DR17 spectral fit. It is 0.01 dex higher than the [Fe/H] value fitted only to Fe lines.}
}
\end{table*}

For KIC\,10001167, we perform separate spectral analysis using either the astrophysically calibrated oscillator strengths of \citet{Slumstrup2019}, or the same lines but the laboratory values of \citet{stetson08}.

With the internal uncertainties presented in Table~\ref{table:abundances}, the results using either are statistically consistent. We adopt only the astrophysically calibrated results based on the line-list of \citet{Slumstrup2019} for further analysis, since it has significantly higher internal precision, less tension between FeI and FeII, and an effective temperature compatible with the IRFM within 3 K (see Sect.~\ref{sec:irfm}). For collective [Fe/H] we use only the [FeI/H] measurements, since the number of FeI lines far surpass the FeII lines, and because the FeII lines of KIC\,10001167 are very weak and therefore sensitive to blending.

We also include effective temperature, metallicity and $\alpha$ enhancement from the APOGEE DR17 \citep{Apogeedr17}, as it has infrared spectra available (apogee\_id 2M19074937+4656118). We find that it is compatible with our analysis within the statistical uncertainties.

Results in Table~\ref{table:abundances} have uncertainties calculated using a standard deviation of the lines, which reflect only line-to-line scatter and not systematic uncertainty.
Therefore, for use in the rest of the analysis we follow the investigations of \citet{Bruntt2010} and adopt a total uncertainty of 0.1dex for [Fe/H] and [$\alpha$/Fe].
\section{Eclipsing binary analysis details}\label{sec:eb-appendix}
\subsection{JKTEBOP}\label{sec:jktebop-details}
To prepare the light curve for JKTEBOP analysis, low-order polynomial fitting of the data near the eclipses is performed to normalize each eclipse.
The light curves are then truncated to keep a minimum amount of data outside of eclipses.
Photometric uncertainty is estimated from the RMS of the phase-folded light curve within the total eclipse.
This is the same procedure as \citep{thomsen2022}, and ensures a homogeneous treatment of the eclipses after trends from reflection, deformation, beaming, and/or activity have been locally removed.
Following this pre-processing, we disable the reflection and deformation approximations in JKTEBOP and treat the stars as spherical.

The PDCSAP light curve has lower apparent photometric noise than the other light curves available to us, since it has been corrected for several known instrumental effects by the mission pipeline.
We also have access to the KASOC light curve \citep{Handberg2014} from the KASOC database \url{kasoc.phys.au.dk}, and the \emph{Kepler} mission pipeline SAP, which in this case has higher photometric noise but better retains out-of-eclipse trends from the binary orbit. We further verify that, after applying the same pre-processing to the KASOC \citep{Handberg2014} light curve and the \emph{Kepler} pipeline SAP, we obtain best-fit results indistinguishable from our PDCSAP analysis. The KEPSEISMIC light curve is unsuitable for eclipse analysis, since it is heavily filtered.

We investigate the \emph{Kepler} target pixel files with Lightkurve \citep{lightkurve2018} and find no sources of contamination near KIC\,10001167 with \emph{Gaia} \emph{G} magnitude below 17. 
Additionally, we find no significant in-system contamination in our investigations of potential spectroscopic contamination (Appendix~\ref{sec:spec-tlight}). 
We therefore treat contamination as negligible for both this, and the PHOEBE, eclipsing binary analysis.

\begin{figure}
\includegraphics[width=\columnwidth, keepaspectratio]{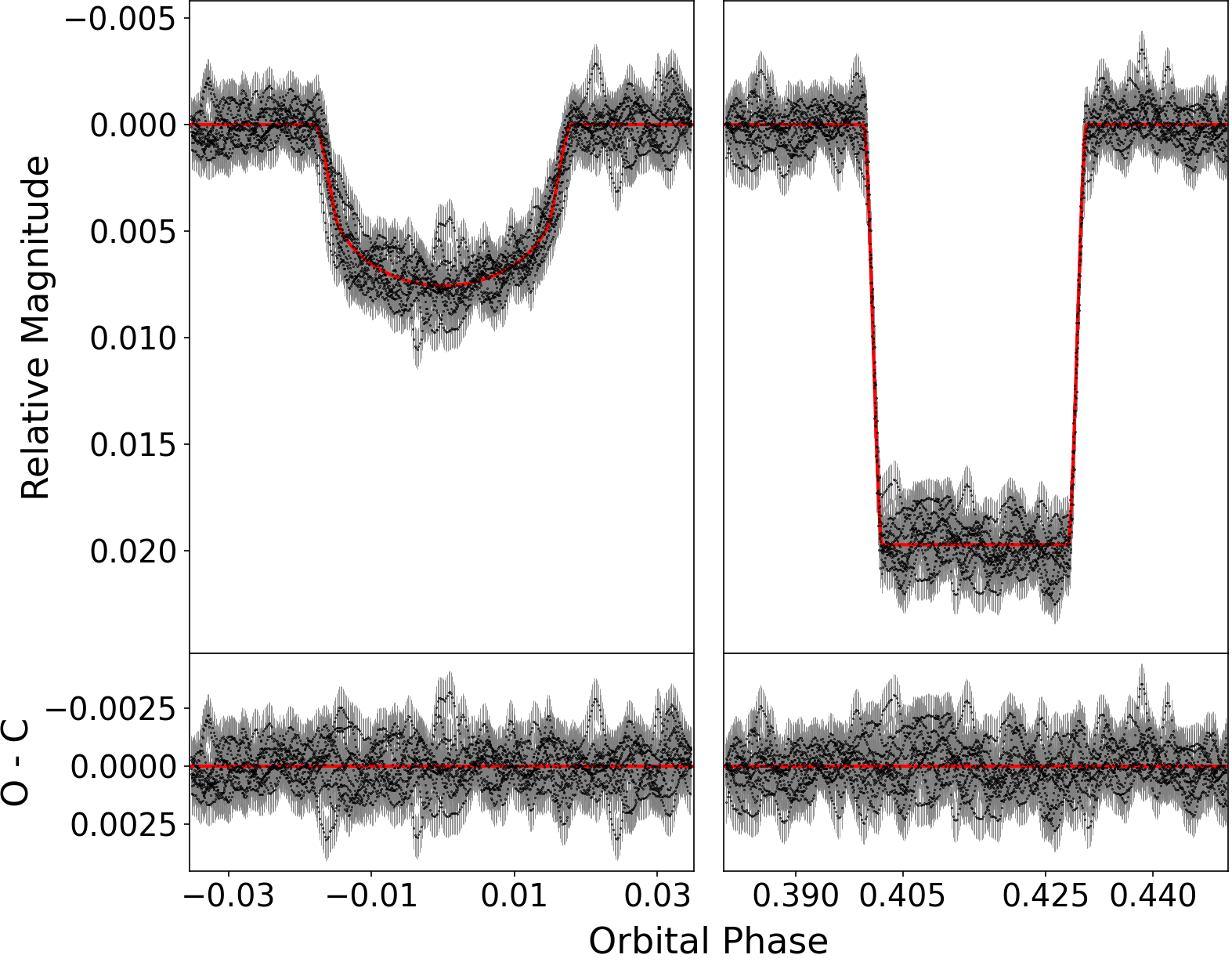}
\caption{Eclipsing binary model fits with JKTEBOP to {\it Kepler} PDCSAP light curve for KIC\,10001167. Observed-Calculated (O-C) residuals are shown below.}
         \label{fig:10001167}%
\end{figure}

In Fig.~\ref{fig:10001167}, the best-fit JKTEBOP light curve model is compared to the \emph{Kepler} PDCSAP light curve.

The limb darkening $h_1$ and $h_2$ coefficients are linearly interpolated from the tables of \citet{Claret2022} for the $K_p$ bandpass, as this parametrization has been found to be superior to other two-parameter limb darkening descriptions when fitting for one coefficient \citep{maxted2023}. \citet{Claret2022} used plane-parallel ATLAS atmosphere models \citep{ck03}. The limb darkening ($h_1$, $h_2$) parametrization of the power-2 law is demonstrated below, as described in \citet{southworth2023}:
\begin{align}
 \frac{F(\mu)}{F(1)} &= 1-c(1-\mu^\alpha),\\
                 h_1 &= \frac{F(0.5)}{F(1)} = 1-c(1-2^{-\alpha}),\\
                 h_2 &= \frac{F(0.5)-F(0)}{F(1)} = c2^{-\alpha}.
\end{align}

\noindent Here, $F(\mu)$ is the flux at position $\mu = \cos\gamma$ along the stellar disc, $\gamma$ is the angle from center, $F(1)$ refers to the central flux, and $c$ and $\alpha$ are the two coefficients for the power law \citep{southworth2023}.

Similar to \citet{thomsen2022}, JKTEBOP was run iteratively to obtain the final limb darkening coefficients using dynamically derived $\log g$'s and $T_{\rm eff}$ for the main sequence star.
In Appendix~\ref{sec:limbdark}, we test the effect of using different formulations of limb or gravity darkening when deriving the radius of the giant, and in Appendix~\ref{sec:eb-atmos} we compare with a light curve model using PHOENIX \citep{husser2013} specific intensities, determining a systematic uncertainty of $\sim 0.7\%$ for the radius of the RG.

\begin{table*}
\centering
\caption{\label{table:EBdata}Properties of KIC\,10001167 obtained from eclipsing binary analysis. The quoted PHOEBE uncertainties are underestimated and should not be used for comparative work (see text).}
\begin{tabular}{lcc}
\hline
\hline
Quantity\tablefootmark{a}    &JKTEBOP   &PHOEBE 2 \\
\hline
$T_{\rm eff,RG}$ (K)& $\sim$& $4804_{-29}^{+40}$ \tablefootmark{a}\\[0.1cm]
SB-ratio, $T_{\rm eff}$-ratio\tablefootmark{b}& $2.937_{-66}^{+69}$ \tablefootmark{a}& $1.29727_{-46}^{+45}$ \tablefootmark{a}\\[0.1cm]
Sum of the fractional radii $r_{\rm MS}+r_{\rm RG}$& $0.1129_{-10}^{+10}$ \tablefootmark{a}& $0.11329_{-22}^{+24}$ \tablefootmark{a}\\[0.1cm]
Ratio of the radii $k$& $0.07642_{-68}^{+70}$ \tablefootmark{a}& $0.07656_{-32}^{+36}$ \tablefootmark{a}\\[0.1cm]
Inclination $i$ ($^{\circ}$)& $87.66_{-16}^{+17}$ \tablefootmark{a}& $87.619_{-46}^{+33}$ \tablefootmark{a}\\[0.1cm]
$e\cos\omega$& $-0.13324_{-18}^{+18}$ \tablefootmark{a}& $-0.133421_{-51}^{+50}$ \tablefootmark{a}\\[0.1cm]
$e\sin\omega$& $-0.0830_{-13}^{+13}$ \tablefootmark{a}& $-0.08280_{-82}^{+87}$ \tablefootmark{a}\\[0.1cm]
Orbital period (days)& $120.39005_{-60}^{+58}$ \tablefootmark{a}& $120.39005_{-22}^{+20}$ \tablefootmark{a}\\[0.1cm]
Reference time $\rm t_{\rm RG}$ (days)& $55028.099_{-14}^{+14}$ \tablefootmark{a}& $55028.1025_{-36}^{+34}$ \tablefootmark{a}\\[0.1cm]
RV semi-amplitude $K_{\rm RG}$  (km/s)& $24.983_{-30}^{+29}$ \tablefootmark{a}& $24.984_{-16}^{+26}$\\[0.1cm]
RV semi-amplitude $K_{\rm MS}$  (km/s)& $27.81_{-11}^{+11}$ \tablefootmark{a}& $27.848_{-52}^{+67}$\\[0.1cm]
System RV $\gamma_{\rm RG}$   (km/s)& $-103.026_{-21}^{+21}$ \tablefootmark{a}& $-103.022_{-19}^{+16}$ \tablefootmark{a,c}\\[0.1cm]
System RV $\gamma_{\rm MS}$   (km/s)& $-102.672_{-90}^{+88}$ \tablefootmark{a}& $-102.724_{-71}^{+68}$ \tablefootmark{a,c}\\[0.1cm]
Semi-major axis $a (R_{\odot})$& $124.17_{-27}^{+27}$& $124.27_{-14}^{+18}$\\[0.1cm]
$a \sin i$ $(R_{\odot})$& $124.07_{-27}^{+27}$ & $124.16_{-14}^{+18}$ \tablefootmark{a}\\[0.1cm]
Eccentricity $e$& $0.15699_{-71}^{+69}$& $0.15703_{-45}^{+45}$\\[0.1cm]
Periastron longitude $\omega$ ($\circ$)& $211.93_{-42}^{+39}$& $211.83_{-28}^{+26}$\\[0.1cm]
Mass-ratio $q = \rm Mass_{\rm MS} / Mass_{\rm RG} $& $0.8984_{-36}^{+37}$& $0.8971_{-20}^{+20}$ \tablefootmark{a}\\[0.1cm]
Mass$_{\rm RG}(M_{\odot})$& $0.9337_{-76}^{+78}$& $0.9362_{-37}^{+50}$\\[0.1cm]
Mass$_{\rm MS}(M_{\odot})$& $0.8388_{-41}^{+42}$& $0.8401_{-20}^{+29}$\\[0.1cm]
Radius$_{\rm RG}(R_{\odot})$& $13.03_{-12}^{+12}$& $13.079_{-36}^{+35}$\\[0.1cm]
Radius$_{\rm MS}(R_{\odot})$& $0.995_{-14}^{+15}$& $1.0014_{-49}^{+52}$\\[0.1cm]
log$g_{\rm RG}$ (cgs)& $2.1786_{-78}^{+77}$& $2.1767_{-24}^{+15}$\\[0.1cm]
log$g_{\rm MS}$ (cgs)& $4.366_{-13}^{+13}$& $4.3617_{-46}^{+39}$\\[0.1cm]
$\rho_{\rm RG} (10^{-3}\rho_{\odot})$& $0.422_{-11}^{+11}$& $0.4189_{-28}^{+23}$\\[0.1cm]
$\rho_{\rm MS} (\rho_{\odot})$& $0.850_{-37}^{+37}$& $0.838_{-13}^{+11}$\\[0.1cm]
Boosting index $b_{\rm RG, Kp}$ & $\sim$ & $5.04_{-19}^{+30}$ \tablefootmark{a}\\ [0.1cm]
ld$_{h1, \rm RG}$& $0.669_{-30}^{+29}$ \tablefootmark{a}& $\sim$\\[0.1cm]
ld$_{h2, \rm RG}$& $0.4187123211$& $\sim$\\[0.1cm]
ld$_{h1, \rm MS}$& $0.7521485458$& $\sim$\\[0.1cm]
ld$_{h2, \rm MS}$& $0.4768031637$& $\sim$\\[0.1cm]
Reduced $\chi^2$-fit& $1.062$& $1.189$\\[0.1cm] 
Reduced $\chi^2$-lc& $1.062$& $1.196$\\[0.1cm]
Reduced $\chi^2$-rvRG& $0.999$& $1.032$\\[0.1cm]
Reduced $\chi^2$-rvMS& $0.763$& $0.831$\\[0.1cm]

\hline
\end{tabular}
\tablefoot{
\tablefoottext{a}{Free parameter during the fit/sampling.}
\tablefoottext{b}{For JKTEBOP, this is the central surface-brightness ratio before limb darkening correction.
For PHOEBE 2, it is the effective temperature ratio.}
\tablefoottext{c}{For PHOEBE, the system RV of the main sequence star is a derived parameter.
The actual fitted parameters are: A shared system RV for the whole system, and an RV offset for the main sequence star.}
}
\end{table*}
\subsection{Radius systematics, Gravity and Limb darkening prescription}\label{sec:limbdark}
Gravity darkening coefficients were fixed at 0.0 for both components during the main JKTEBOP analysis.
Using gravity darkening coefficients from \citet{Claret2011}, we found no measurable differences in the parameters.

The limb darkening formulations described in this section can all be found in \citet{southworth2023}.
When using the re-parametrized power-2 limb darkening law, but $h_2$ coefficient instead interpolated from the tables of \citet{Claret2023}, which used spherical PHOENIX-COND atmosphere models \citep{husser2013} with solar abundances, the RG radius is increased by $0.2\%$ ($0.2\sigma$).
When using the four-parameter limb darkening law, and four coefficients fixed to interpolated values from \citet{Claret2023}, the radius is lowered by $0.06\%$ ($0.07\sigma$).
We also tested several different two-parameter formulations keeping both coefficients fixed;
1. power-2 law with ATLAS-derived coefficients from \citet{Claret2022};
2. quadratic law with ATLAS-derived coefficients from \citet{Claret2011};
3. quadratic law with ATLAS-derived coefficients from \citet{Sing2010}.
The maximum deviation in RG radius, in-between these and when comparing each with the one with h1 free (and $h_2$ from \citet{Claret2022}), was $0.3\%$ ($0.3\sigma$).
When increasing the effective temperature of the giant by 100K (and the MS star by 130K) while fitting $h_1$, the measured radius decreased by $0.05\%$ ($0.05\sigma$).
When increasing the RG temperature by 200K (and MS by 260K) while fitting $h_1$, the measured radius decreased by $0.09\%$ ($0.1\sigma$).
We note that the four-parameter law with pure-interpolated coefficients, which is the law best reproducing theoretical atmosphere models \citep{Claret2023}, produces a radius that is almost an exact match with our result determined using power-2 law with $h_1$ free (deviation $0.06\%$), despite using a different atmosphere model (PHOENIX instead of ATLAS).
In summary, the obtained radius is robust regardless of the exact limb darkening law used.
Variations caused by the assumed limb darkening law (at maximum $0.3\%$) are insignificant compared with the uncertainty on the radius, which is $\sim 0.9\%$.
\subsection{Radius systematics, Atmosphere and specific intensity}\label{sec:eb-atmos}
As implemented, all the limb darkening prescriptions tested in Sect.~\ref{sec:limbdark} assume that the stellar atmosphere has a hard cut-off at the edge of the limb.
This cut-off is taken as the point where the gradient of the specific intensity profile of the star is at maximum.
We tested the validity of this approximation using the program \emph{ellc} \citep{maxted2016}, which allows for direct input of specific intensities.
For the RG, we used specific intensities from PHOENIX-COND atmosphere models \citep{husser2013}, interpolated to the effective temperature, metallicity and dynamical $\log g$, and integrated over the \emph{Kepler} passband.
Then, we generated a synthetic light curve with the complete set of specific intensities, and re-fitted the stellar radii, surface brightness ratio and inclination with a model using the aforementioned specific intensity cut-off.
The re-fitted radii were $\sim0.3\%$ higher than the input radii.
This demonstrates the direct influence that our assumption of a hard stellar surface has under perfect conditions.

After, we performed the same simulation, but for the re-fit model we utilized either the four-parameter limb darkening law with the same coefficients as Sect.~\ref{sec:limbdark}, or the power-2 law.
In both cases, we find that the re-fitted RG radius decreases by $0.7\%$ (and MS increases by $1.7\%$).
It is also evident, from comparing the specific intensities, that both the limb darkening laws poorly reproduce the atmosphere model profile.
We take the maximum deviation reported, from this and Sect~\ref{sec:limbdark}, as the combined systematic uncertainty on the dynamical radius of the RG.

\subsection{Light travel time}\label{sec:ltte}
In our JKTEBOP and PHOEBE analyses, light travel time effects have not been accounted for.
The differential effect due to the orbit is negligible, but the constant system radial velocity has a significant effect on the perceived orbital period.
We tested this with JKTEBOP by performing a linear correction of the time-stamps of all data (light curve and radial velocities) using an approximate system velocity of $-103$ km/s, to account for the increasing distance between the target and the solar system.
The measured orbital period is then decreased to $120.34868$ days, a change of $0.03\%$ ($69\sigma$). This decreases the orbital semi-major axis in turn by $0.03\%$ ($0.2\sigma$), the mass of the RG by $0.03\%$ ($0.04\sigma$), and the radius of the RG by $0.03\%$ ($0.04\sigma$). Since the interest of the analysis is the fundamental stellar parameters, this effect can be safely ignored for KIC\,10001167.
\subsection{PHOEBE 2}\label{sec:phoebe-details}
For subsequent analysis with PHOEBE 2 \citep{conroy2020}, we start from the unfiltered KASOC light curve. The PDCSAP light curve, while having lower noise due to the cotrending basis-vector corrections of the \emph{Kepler} data analysis pipeline, suffers from over-fitting of the eclipsing binary signal from those same corrections. Since they are multiplicative, it is possible to correct for them by normalizing the eclipses (as we have done and demonstrated in Appendix~\ref{sec:jktebop-details}). However, for PHOEBE we wish to model both the eclipses and out-of-eclipse signal of binarity, which makes the PDCSAP light curve unsuitable. The \emph{Kepler} pipeline SAP light curve would be a reasonable choice, but we decided to use the KASOC light curve, which is also based on simple aperture photometry, because it uses a larger aperture which reduces the impact of spacecraft motion on the light curve. We perform a custom, iterative filtering of the KASOC light curve, inspired by \citet{Handberg2014}. Here, we start with a window size much longer than the orbit for our non-phased long-trend filter, and incrementally reduce it as the transit is captured by a phase-folded filtering.
This filtering is advantageous since it retains the complete (repeating) eclipsing binary signal both inside and outside of eclipses with limited over-fitting.

For the MCMC sampling with \emph{emcee}, we use uninformative uniform priors around the JKTEBOP results, with boundaries at least four times the measured JKTEBOP uncertainties, and extend them if any parameter converged towards the edges.

The uncertainties we obtain from our MCMC sampling are heavily underestimated, and should not be used for comparing with other analyses.
The sampling assumes that no residual signal remains in the light curve besides the binary orbit signal, while the residuals for KIC\,10001167 are clearly dominated by asteroseismic signal.
We verify this by performing an independent MCMC sampling with the JKTEBOP light curve modelling and recovering similarly underestimated uncertainties.
Due to computational expense, it is not possible to perform as detailed an investigation of the uncertainties as we perform with the JKTEBOP analysis.
We therefore use the latter as the baseline result for comparison with asteroseismic, and photometric, inference.

\begin{figure}
\includegraphics[width=\columnwidth, keepaspectratio]{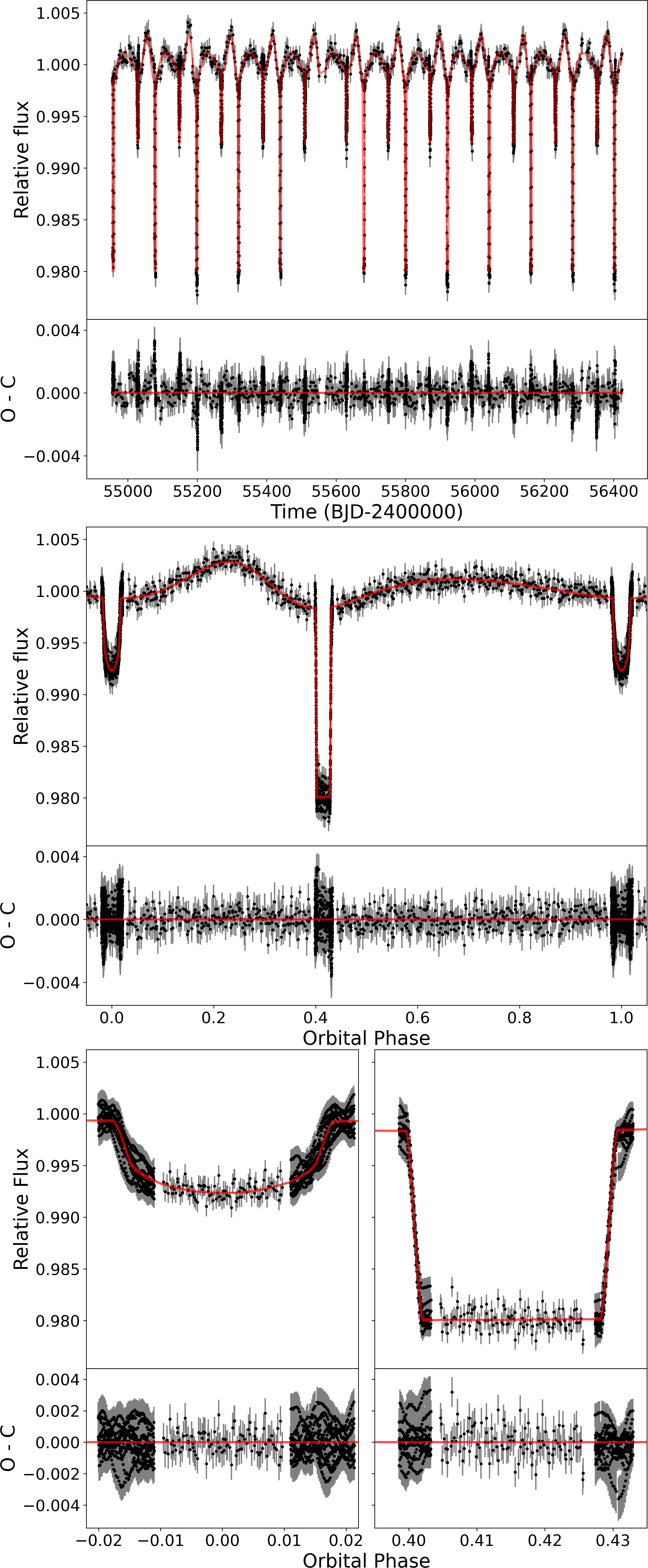}
\caption{Eclipsing binary model fits with PHOEBE 2 to independently filtered KASOC light curve for KIC\,10001167. Top panels illustrate the binned light curve (black points) and model (red) with time on the x-axis. Middle panels show the same, but with orbital phase on the x-axis. The bottom panels show zoomed-in views of just the two eclipses (phase-folded). The data have been binned in time, in three different ways depending on phase. See the text for details. The gaps between different binnings is a consequence of time-averaging within the bins, and a hard cut to avoid data overlap between them.}
         \label{fig:phoebe_10001167}%
\end{figure}
\section{Spectroscopic test for a potential unresolved third star}\label{sec:spec-tlight}
Using the FIES spectroscopy, we explored the possibility of stellar light contamination from an unresolved in-system third companion, to rule out biases on the dynamical RG radius therefrom.
It was performed in a similar way to \citet{brogaard2022}.
We subtract the separated component spectra from all observed spectra that have RVs for both the RG and the MS star, and then average them.
Since the system velocity is found to be constant, a potential third stellar companion would have to be on a wide orbit and therefore have almost constant RV.
If it is gravitationally bound, it should have low RV relative to the system.
With the combined spectrum, we make a broadening function profile using the MS template spectrum.
Here, we find no significant single peak, only spurious peaks $1.5$ times the noise exactly within the $\pm 25$ km/s region, a clear leftover systematic from the RG spectral separation.

We then inject an artificial third star, by rotationally broadening the MS template with $v \sin i = 5$ km/s, with noise calculated from the injection light ratio and the SNR of each spectrum.
We adjust the light contribution from the third signal until the broadening function signal is 2 times the $\pm 25$km/s noise (3 times the noise outside), which occurs when the injected luminosity ratio is $0.5\%$.
We set this as the detection threshold for a third companion.
Assuming a luminosity of $70 \rm L_{\odot}$ for the RG, this puts an upper limit of $\sim 0.35 \rm L_{\odot}$ for an undetected third star.
Statistically, such a star would be on the main sequence, meaning spectral type K3V or cooler\footnote{From \url{https://www.pas.rochester.edu/~emamajek/EEM_dwarf_UBVIJHK_colors_Teff.txt}, see \citet{pecaut2013}.}. It is unlikely that its brightness would be exactly at the detection limit. A contaminant with brightness far below the limit would not bias the light curve radius measurement.
\section{Parallax, photometry, IRFM, and kinematics}\label{sec:parallax-phot-irfm}
Gaia Data Release 3 (DR3) offers parallax and photometry for KIC\,10001167 \citep{Gaia2016, gaia2022}.
Table~\ref{table:astrometry} shows the astrometric parameters from \emph{Gaia} DR3, including an additional uncertainty estimate due to the binary orbit, which we derive in Appendix~\ref{sec:parallax-error}.
Table~\ref{table:phot-rg} includes photometric measurements of the RG derived using the DR3 parallax and either single-passband photometry with \emph{Gaia} DR3 or 2MASS \citep{2MASS2006} or the infra-red flux method (IRFM). It also includes MS star photometric contributions estimated in the IRFM using the eclipsing binary radius ratio (JKTEBOP) and effective temperature ratio (PHOEBE), in combination with DR3 photometry in the G, BP, and RP bands.
\subsection{Parallax uncertainty}\label{sec:parallax-error}
Assuming that the parallax-derived distance of Sect.~\ref{sec:distance} is accurate, the orbital motion on the sky is expected to be significant in comparison to the parallax, with semi-major axis of 0.667 mas if the system is aligned for \emph{maximum} movement on the sky.
It could therefore bias the parallax measurement.
We use the method mentioned in \citet{brogaard2022} and detailed in \citet[Appendix C]{rappaport2022} to estimate an additional uncertainty contribution from this.

Summarizing, we start by first assuming that the 35 astrometric transits are all instantaneously measured, and equi-spaced in time between the \emph{Gaia} DR3 start and end dates of observation (2014/07/25 to 2017/05/28).
To simplify the computation, we assume that the inclination is $i=90^{\circ}$ such that the orbital movement happens only along a line on the sky, and we assume that the orbit is aligned for \emph{maximum} variation on the sky.
Then, we calculate the centre of light distance from the centre of mass using the eclipsing binary masses, luminosity ratio, semi-major axis, period, eccentricity, and periastron time.
This is converted into a projected distance on the sky, in milli-arcseconds, for each simulated astrometric transit.
From this, we calculate a standard deviation, and convert it into an approximate standard error by dividing with the square-root of the number of astrometric transits minus the number of astrometric fitting parameters (35-5).
With this, we obtain a standard error of 0.040 mas, despite having a maximum orbital semi-major axis of 0.667 mas.
These results, along with the \emph{Gaia} DR3 astrometric parameters, can be found in Table~\ref{table:astrometry}.

\subsection{Extinction}\label{sec:extinction}
We estimated the extinction using three different 3D dust maps: 1. \emph{Bayestar19}, available through the python package \emph{dustmaps} \citep{green2018, green2019}.
2. The map by \citet{lallement2019}.
3. The updated map by \citet{lallement2022}.

The \emph{Bayestar19} map was converted to \emph{E(B-V)} using the two example relations on their documentation website\footnote{\url{http://argonaut.skymaps.info/usage}} (either \emph{E(B-V)}$\rm =0.884 \times (Bayestar19)$ or \emph{E(B-V)}$\rm = 0.996\times(Bayestar19)$.
Then, they were converted to passband-specific extinction coefficients $\rm A_X$ using the color-relations of \citet[Table B1]{casagrande2021} (specifically, the EDR3 FSF relations).
\emph{(BP-RP)}$_0$ was derived iteratively, starting with \emph{(BP-RP)}-\emph{E(B-V)} as initial guess.

From the \citet{lallement2019} and \citet{lallement2022} maps, extinction at 550nm ($\rm A_0$) was obtained.
To convert this to passband-specific extinction, we utilized the \emph{Gaia} EDR3 auxiliary data files\footnote{\url{https://www.cosmos.esa.int/web/gaia/edr3-extinction-law}}.
Here, we specifically made use of the \emph{(BP-RP)} color-relations for stars at the top of the HR diagram (Giants).

When we estimate extinction using the three different maps, while neglecting that KIC\,10001167 is a binary, we find general agreement on a very low amount of extinction, but otherwise it is quite uncertain with \emph{E(B-V)} between 0.03 and 0.04.

Due to this, for the IRFM and luminosity measurements we adopt a simple estimate of $\rm \emph{E(B-V)}=0.0350 \pm 0.0105$ which is in between the different measurements.
Here, the uncertainty is conservatively taken as $30\%$ of the value, which means that all three maps are well within $1\sigma$.
This large uncertainty ensures that we do not bias the final result due to selection of a specific map.
At the same time however, since the reddening is small, it is not going to inflate the propagated uncertainties significantly.
\begin{table*}
\centering
\caption{\label{table:phot-rg}IRFM measurements, photometry, and reddening of the stellar components in KIC\,10001167, including magnitude corrections to account for the companion in the photometry. The IRFM used the temperature ratio $\rm T_{MS}/ T_{RG} = 1.2973 \pm 0.0076$ and radius ratio $\rm R_{MS} / R_{RG} = 0.07642 \pm 0.00069$ from the eclipsing binary analysis. A reddening estimate of $\rm \emph{E(B-V)} = 0.0350\pm0.0105$ is used, a conservative value that puts all three reddening maps discussed in the text within $1\sigma$. The distance used is $866 \pm 35$pc.
To obtain the single-passband estimates of luminosity and radius of the RG, the following parameters are also used for bolometric corrections: $\rm [Fe/H] = -0.68 \pm 0.1$ dex, $\rm T_{eff, RG} = 4625 \pm 29 (normal) \pm 30 (uniform)$ K, with \emph{Gaia} DR3 FSF extinction coefficients from \citet{casagrande2021} ([Table B1]). SB-radius and SB-luminosity refers to the Stefan-Boltzmann radius and luminosity.}
        \begin{tabular}{lcc}
\hline
\hline
IRFM & & \\
\hline
Angular diameter, RG (mas)  & \multicolumn{2}{c}{$\rm 0.1376\pm0.0017$ (stat) $\pm 0.0003 $ (syst)} \\
Angular diameter, MS (mas)  & \multicolumn{2}{c}{$\rm 0.01055\pm0.00022$ (stat) $\pm 0.0003$ (syst)} \\
$\rm T_{eff, RG}$ (K) & \multicolumn{2}{c}{$\rm 4625 \pm 29$ (stat) $\pm 30 $ (syst)}\\
$\rm T_{eff, MS}$ (K) & \multicolumn{2}{c}{$\rm 6031 \pm 108$ (stat) $\pm 30 $ (syst)}\\
Radius $R_{\rm RG}$ ($\rm R_{\odot}$) & \multicolumn{2}{c}{$12.82 \pm 0.30$ (stat) $\pm 0.24$ (syst)}\\ 
Radius $R_{\rm MS}$ ($\rm R_{\odot}$) & \multicolumn{2}{c}{$0.983 \pm 0.47$}\\ 
SB-luminosity, RG ($\rm L_{\odot}$) & \multicolumn{2}{c}{$67.7 \pm6.1$}\\
SB-luminosity, MS ($\rm L_{\odot}$) & \multicolumn{2}{c}{$1.15 \pm0.14$}\\
\hline
Companion correction, photometry & RG (dex) & MS (dex) \\
\hline
Gaia DR3 $BP$ & $0.0259 (12)$ & $4.022 (54)$ \\
Gaia DR3 $G$  & $0.0206 (8)$ & $4.275 (45)$ \\
Gaia DR3 $RP$ & $0.0166 (6)$ & $4.516 (43)$\\
2MASS $J$ & $0.0117 (3) $ & $4.903 (32)$ \\
2MASS $H$ & $0.0089 (2)$ & $5.204 (24)$\\
2MASS $Ks$ & $0.0085 (2)$  & $5.256 (23)$\\
\hline
Photometry & RG (mag) & MS (mag)\\
\hline
Gaia DR3 $BP$ & $10.6213 (14)$ & $14.617 (54)$ \\
Gaia DR3 $G$  & $10.06685 (82)$ & $14.321 (45)$\\
Gaia DR3 $RP$ & $9.36093 (72)$ & $13.860 (43)$ \\
Gaia DR3 \emph{(BP-RP)} & $1.2604 (16)$ & $0.757 (69)$\\
2MASS $J$ & $8.407 (23)$ & $13.298 (39)$\\
2MASS $H$ & $7.858 (36)$ & $13.053 (43)$\\
2MASS $Ks$ & $7.756 (23)$ & $13.003 (33)$ \\
\hline
Reddening \& extinction, RG & \multicolumn{2}{c}{dex} \\
\hline
Assumed \emph{E(B-V)} & \multicolumn{2}{c}{0.0350 (105)} \\
\emph{(BP-RP)}$_0$ & \multicolumn{2}{c}{1.223 (11)} \\
$\rm A_{BP}$ & \multicolumn{2}{c}{0.094 (28)} \\
$\rm A_G$ & \multicolumn{2}{c}{0.074 (22)} \\
$\rm A_{RP}$ & \multicolumn{2}{c}{0.057 (17)} \\
$\rm A_{J}$ & \multicolumn{2}{c}{0.0252 (75)} \\
$\rm A_{H}$ & \multicolumn{2}{c}{0.0159 (48)} \\
$\rm A_{Ks}$ & \multicolumn{2}{c}{0.0107 (32)} \\
\hline
Bolometric corrections, RG & \multicolumn{2}{c}{dex} \\
\hline
$\rm BC_{BP}$ & \multicolumn{2}{c}{-0.651 (18)} \\
$\rm BC_G$ & \multicolumn{2}{c}{-0.128 (10)} \\
$\rm BC_{RP}$ & \multicolumn{2}{c}{0.5498 (39)} \\
$\rm BC_{J}$ & \multicolumn{2}{c}{1.4461 (79)} \\
$\rm BC_{H}$ & \multicolumn{2}{c}{2.019 (18)} \\
$\rm BC_{Ks}$ & \multicolumn{2}{c}{2.145 (20)} \\
\hline
Bolometric luminosity | SB-radius, RG & $\rm L_{\odot}$ & $\rm R_{\odot}$ \\
\hline
$\rm L_{bolo, BP}$ ~~|~~ $\rm R_{BP}$& 66.1 (57)  &  12.66 (60)\\
$\rm L_{bolo, G}$ ~~~|~~ $\rm R_{G}$& 66.8 (56)  &  12.73 (58)\\
$\rm L_{bolo, RP}$ ~~|~~ $\rm R_{RP}$& 67.5 (56)  &  12.80 (56)\\
$\rm L_{bolo, J}$ ~~~~|~~ $\rm R_{J}$& 69.2 (58)  &  12.95 (56)\\
$\rm L_{bolo, H}$ ~~~|~~ $\rm R_{H}$& 67.1 (60)  &  12.76 (56)\\
$\rm L_{bolo, Ks}$ ~~|~~ $\rm R_{Ks}$& 65.3 (56)  &  12.59 (53)\\
\hline
\end{tabular}
\end{table*}
\subsection{Photometry and Infra-red Flux Method}\label{sec:irfm}
We use the IRFM to simultaneously estimate the effective temperature and angular diameter of the RG.
The IRFM implementation is described in \cite{casagrande2021}, which is based on \emph{Gaia} and 2MASS photometry and has been extensively validated against interferometric and other benchmark stars.
We fix [Fe/H]=$-0.68$ (APOGEE DR17), $\log g = 2.18$ and $E(B-V)=0.035$ (from Appendix~\ref{sec:extinction}) to obtain an initial $T_{\rm eff}=4638$~K and $\theta=0.1379$~mas for the RG.
These values, together with the ratios of effective temperatures and radii from Table \ref{table:EBdata}, are used to correct \emph{Gaia} and 2MASS photometry for the flux contribution of the MS star before running the IRFM again.
We converge to $T_{\rm eff}=4625$~K and $\theta=0.1376$~mas in one iteration, these values being identical regardless of adopting JKTEBOP or PHOEBE 2 ratios.
Uncertainties in the flux contribution of the MS star are derived with 10,000 Monte Carlo realizations assuming normal errors for the effective temperature and radius ratios (Table~\ref{table:EBdata}), metallicity (0.1~dex), gravity (0.1~dex), reddening (0.0105), 2 and 1 percent for the initial effective temperature and radius of the red giant.

It can be appreciated from Table \ref{table:phot-rg} that the photometric contribution of the MS star is at most a few hundredths of a magnitude, with negligible uncertainties in all cases.
To estimate final uncertainties on the effective temperature and angular size of the RG we run another 10,000 Monte Carlo realizations drawing again from a normal distribution in the adopted values of metallicity, reddening and photometry.
These statistical uncertainties are provided in Table~\ref{table:phot-rg} along with systematic ones for the IRFM and parallax. Systematic errors within the IRFM take into account the uncertainty of the zero point of the adopted $T_{\rm eff}$ scale, inflated by a further 10K if we were to adopt a different reddening law \citep[COD instead of FSF, see][for further details]{casagrande2021}.

Using the angular size of the RG with the \emph{Gaia} DR3 distance measurement, we obtain a radius of $12.82\pm 0.30$ (stat) $\pm 0.24$ (syst) $\rm R_{\odot}$. Combining effective temperature with photospheric radius, we measure a luminosity of $67.7 \pm 6.1 \rm L_{\odot}$. We report the IRFM results, along with single-passband bolometric luminosities assuming bolometric corrections from \emph{bolometric-corrections} (\url{https://github.com/casaluca/bolometric-corrections}) software \citep[e.g. ][and references therein]{casagrande2018b}, in Table~\ref{table:phot-rg}. The bolometric corrections are based on MARCS model fluxes \citep{Gustafsson2008}, whereas the IRFM we describe in this section relies on model fluxes from \cite{ck03}.

Table~\ref{table:phot-rg} includes all the binary-corrected photometry, IRFM-based measurements, as well as luminosity and radius from single-passband measurements with all the \emph{Gaia} DR3 and 2MASS passbands. The main contribution to the uncertainty in both approaches (IRFM vs. single-passband) is the parallax.
\begin{table}
\centering
\caption{\label{table:astrometry}Astrometry and kinematics of KIC\,10001167. See text for description.}
\begin{tabular}{lc}
\hline
\hline
Astrometry & Value \\
\hline
Gaia DR3 values: & \\
Parallax (mas) & 1.135 \\
Parallax zero-point (mas)\tablefootmark{a} & -0.019 \\
Parallax error (mas) & 0.018 \\
Parallax error, inflated (mas)\tablefootmark{b} & 0.022 \\
Renormalized unit weight error, RUWE & 1.349 \\
\hline
This work: & \\
Semi-major axis $a$(col) (mas)\tablefootmark{c} & 0.6669\\
Standard-deviation(col) (mas)\tablefootmark{c} & 0.22\\
Error(col) (mas)\tablefootmark{c} & 0.040 \\
Combined error (mas) & 0.046\\
\hline
Distance (pc) & $866_{-33}^{+36}$ \\
\hline
\hline
Galactic orbital parameters\tablefootmark{d} & Value \\
\hline
U (km/s) & $-23_{-1}^{+1}$ \\
V (km/s) & $-104_{-4}^{+5}$\\
W (km/s) & $-4_{-1}^{+1}$\\
$v_{\rm perpendicular}$ (km/s) & $23.51_{-0.4}^{+1}$ \\
R (kpc) & $8.07$ \\
$\phi$ (rad) & $3.04$ \\
z (kpc) & $0.27$ \\
$x_{\rm HC}$ (kpc) & $0.18$ \\
$y_{\rm HC}$ (kpc) & $0.82$ \\
$z_{\rm HC}$ (kpc) & $0.26$ \\
Eccentricity & $0.42_{-0.02}^{+0.02}$ \\
$R_{\rm guide}$ (kpc) & $4.8_{-0.2}^{+0.2}$ \\
$Z_{\textup{max}}$ (pc) & $35_{-0.4}^{+0.8}$ \\
$J_{R}$ (kpc km s$^{-1}$) & $172_{-14}^{+15}$ \\
$J_{Z}$ (kpc km s$^{-1}$) & $2.4_{-0.1}^{+0.1}$ \\
$L_{Z}$ (kpc km s$^{-1}$) & $1059_{-40}^{+40}$ \\
Orbital energy $E$ (km$^{2}$ s$^{-2}$/1e5) & $-1.76_{-0.06}^{+0.06}$ \\
$L_z / L_c$ & 0.82 \\
\hline
\end{tabular}
\tablefoot{
\tablefoottext{a}{ from \citet{lindegren2021}.}
\tablefoottext{b}{ Using \citep{Apellaniz2021}.}
\tablefoottext{c}{ "col" meaning "centre of light".}
\tablefoottext{d}{ Using radial velocity of the red giant from JKTEBOP, Table~\ref{table:EBdata}.}
}
\end{table}
\subsection{Distance}\label{sec:distance}
The uncorrected parallax is $1.135 \pm 0.018$mas.
The parallax error is increased using the re-scaling of the raw parallax error suggested by \citet{Apellaniz2021}, as well as with the additional parallax uncertainty estimate from the binary orbital motion that we find in Appendix~\ref{sec:parallax-error}.
The parallax zero-point is evaluated to be $-0.019 \mathrm{mas}$ using the python package \emph{gaiadr3-zeropoint} \citep{lindegren2021} and subsequently corrected for.

Excluding (including) the extra parallax uncertainty from the orbit, we measure a distance of $866_{-17}^{+18}$ ($_{-33}^{+36}$) pc when the parallax zero-point is corrected for. Without zero-point correction, we obtain $881$pc.
Both the geometric ($866\pm 13$pc) and photogeometric ($862 \pm 11$pc) distances from \citet{BailerJones21} are fully compatible with the zero-point corrected distance measurement we provide here.
We propagate our additional uncertainty estimate from orbital motion for subsequent radius and luminosity inference. However, we use the final difference between the two uncertainty estimates as our reported systematic uncertainty estimate.
\section{Asteroseismic details}\label{sec:asteroseismology_appendix}
\subsection{MESA grid}\label{sec:mesa}
The MESA (version n. 11701) grid used for asteroseismic inference spans a wide range of [Fe/H], $\rm [\alpha/Fe]$ and helium abundance ($Y$) values. [Fe/H] ranges from -2.0 to 0.25 in steps of 0.25; $Y$ ranges from 0.25 to 0.34 in steps of 0.03. We also consider four values of $\alpha$-enhancement, $\rm [\alpha/Fe]$=-0.2,0.0,+0.2,+0.4. The reference solar mixture is the one from \citet{Asplund2009}. 
For each choice of [Fe/H], $Y$ and $\rm [\alpha/Fe]$ values, we calculate models with masses between 0.6 and 1.8 $\rm M_{\odot}$, with a step of 0.05  $\rm M_{\odot}$.
We adopt the atmosphere description by \citet{KrishnaSwamy1966} and other choices in terms of equation of state, opacities, and mixing are as in \citet{Miglio2021}. 
Finally, we use the mixing-length-theory (MLT) convection scheme in the formulation provided by \citet{cox1968}, where we set the $\rm \alpha_{MLT}$ parameter to 2.2902. This value is the result of a standard solar-model calibration process.
Each model is evolved from the pre-main-sequence to the first thermal pulse.
Adiabatic oscillation frequencies of angular degree l=0 and 2 are computed using the code GYRE  \citep[][and references therein]{townsend2013}.  For RGB stars with $\log_{10}(\langle\rho\rangle/\rho_{\rm centre})$ larger than 8.0 we compute non-radial modes using the approximation presented in \citet{Ong2020ApJ}. As described by Tailo et al. (in preparation), this ensures that the frequencies of quadrupolar p-dominated modes in RGB stars can be computed accurately (to better than 0.05\% at $L\sim 65  \mathrm {L_\odot}$) and in a computationally efficient manner. In low-mass stars, this threshold is met at $L\gtrsim 30  \mathrm {L_\odot}$, hence it is fully justified for KIC\,10001167. 
\subsection{Frequency extraction}\label{sec:frequency_extraction}
\begin{figure}
    \includegraphics[width=\columnwidth, keepaspectratio]{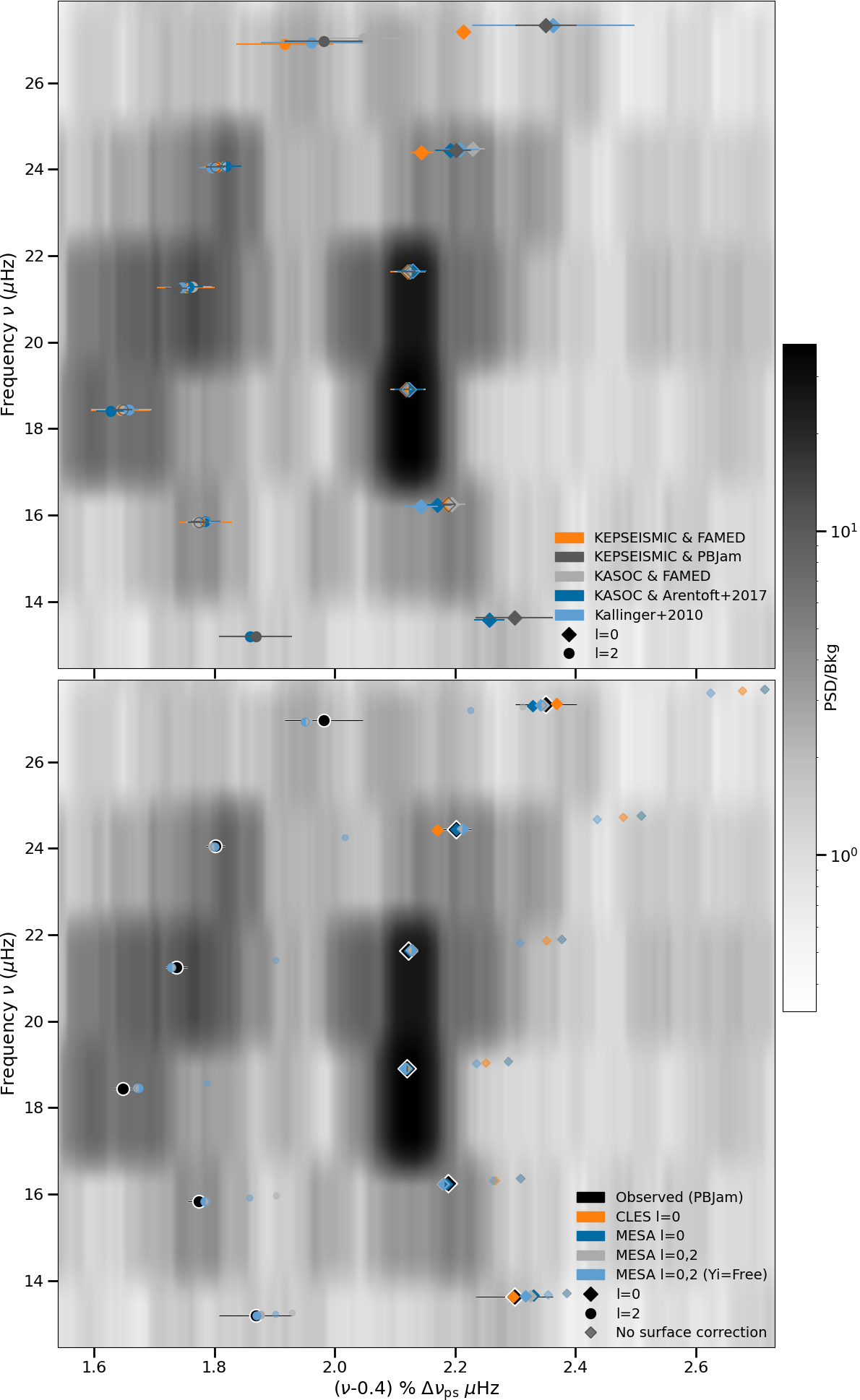}
    \caption{Echelle diagrams for the asteroseismology of KIC\,10001167. X-axis shows frequency $\nu$ modulo 2.731 $\mu$Hz (offset), and Y-axis shows frequency. Heatmap data is uniformly smoothed with window=$0.075\mu$Hz. Marker filling has been half/fully removed for some of the overlapping points. Top panel: l=0 and l=2 frequencies observed with five different methods described in Section~\ref{sec:freq_numax}. Only frequencies found with at least two methods are shown. Bottom panel: Best fit theoretical frequencies vs. a single set of observed frequencies (KEPSEISMIC+PBJam) that were used as observational constraints for the inference.}
    \label{fig:echelle}
\end{figure}

Fig.~\ref{fig:echelle} shows the agreement between all the observed frequencies of this work, as well as \citep{kallinger2019}. Table~\ref{table:RGseis} gives all frequencies observed by at least two pipelines.
\begin{table*}
\centering
\caption{\label{table:RGseis}Top: Asteroseismic global parameters of the RG in KIC\,10001167. The $\rm \nu_{max}$ measurements are from the following sources (see Sect.~\ref{sec:freq_numax} for descriptions of the first three): i. KEPSEISMIC+FAMED. ii. KASOC+FAMED. iii. KASOC+\citep{Arentoft17} method. iv. \citep{montalban2021}. v. \citep{yu2018}. vi. \citep{Gaulme2016}. vii. \citep{kallinger2010}. viii.  \citep{mosser2009}. ix. \citep{hekker2010}. x. \citep{huber2009}. Bottom: Observed asteroseismic frequencies of the red giant. See text for details on the different methods. Only showing frequencies detected by at least two pipelines.}
\begin{tabular}{lccccc}
\hline
\hline
\multicolumn{2}{l}{Quantity} & \multicolumn{4}{c}{Value} \\
\hline
\multicolumn{2}{l}{$\Delta\nu_{\rm ps}$ ($\mu$Hz)}              &\multicolumn{4}{c}{$2.731 (13)$}\\
\multicolumn{2}{l}{$\Delta\nu_0$ ($\mu$Hz)}                                     & \multicolumn{4}{c}{$2.714 (17)$}\\
\multicolumn{2}{l}{$f_{\Delta\nu}$\tablefootmark{a}}         &\multicolumn{4}{c}{$0.95863$}\\
\multicolumn{2}{l}{$\nu_{\rm max}$ ($\mu$Hz)}    & \multicolumn{1}{l}{i. ~~$20.103 (366)$} & \multicolumn{1}{l}{ii. ~$19.784 (158)$} & \multicolumn{1}{l}{iii. \,$19.65 (10)$} & \multicolumn{1}{l}{iv. ~\>$19.68 (33)$} \\
& & \multicolumn{1}{l}{v. ~$19.93 (44)$} & \multicolumn{1}{l}{vi. \,$19.90 (7)$} & \multicolumn{1}{l}{vii. $20.038 (127)$} & \multicolumn{1}{l}{viii. $19.70 (35)$} \\
& & \multicolumn{1}{l}{ix. $19.68 (18)$}     & \multicolumn{1}{l}{x. ~$19.92 (45)$} & &                                    \\
\hline
\hline
(n, l) & A+FAMED & B+FAMED & C+\citep{Arentoft17} & A+PBJam & D \\
\hline
(3, 2)      &             &             & $13.1840 (274)$ SN 1.3  & $13.197 (61)$&  \\
(4, 0)      &             &             & $13.5806 (256)$ SN 2.7  & $13.627 (64)$&  \\
(4, 1)      &             &             & $15.1617 (247)$ SN 6.5  & &  \\
(4, 2)      & $15.8455 (449)$  & $15.8256 (181)$   & $15.8403 (246)$ SN 7.5 & $15.834 (17)$& $15.840 (25)$ \\
(5, 0)      & $16.2474 (222)$   & $16.2496 (227)$   & $16.2251 (246)$ SN 7.1 & $16.249 (12)$& $16.198 (28)$ \\
(5, 1)      & $17.6567 (603)$   & $17.6706 (451)$   & $17.6670 (242)$ SN 19.5 & &  \\
(5, 2)      & $18.4370 (491)$   & $18.4317 (509)$   & $18.4139 (243)$ SN 14.5 & $18.440 (16)$& $18.444 (17)$ \\
(6, 0)      & $18.9109 (263)$   & $18.9087 (288)$   & $18.9093 (257)$ SN 42.1 & $18.912 (5)$& $18.909 (5)$   \\
(6, 1)      & $20.4459 (465)$   & $20.4267 (419)$   & $20.4361 (242)$ SN 25.6 & &  \\
(6, 2)      & $21.2769 (482)$   & $21.2808 (316)$   & $21.2741 (242)$ SN 25.2 & $21.261 (18)$& $21.262 (15)$ \\
(7, 0)      & $21.6452 (285)$   & $21.6417 (267)$   & $21.6452 (241)$ SN 38.5 & $21.647 (10)$& $21.647 (10)$ \\
(7, 1)      & $23.0965 (216)$   & $23.1025 (225)$   & $23.0943 (242)$ SN 20.9 & &  \\
(7, 2)      & $24.0602 (228)$   & $24.0665 (218)$   & $24.0683 (242)$ SN 19.7 & $24.058 (16)$& $24.044 (22)$ \\
(8, 0)      & $24.4011 (195)$   & $24.4768 (202)$   & $24.4396 (243)$ SN 17.6 & $24.458 (25)$& $24.456 (32)$ \\
(8, 1)      & $26.0151 (228)$   & $26.0034 (183)$   &                  & &  \\
(8, 2)      & $26.9052 (806)$   & $27.0274 (710)$   &                  & $26.970 (65)$& $26.941 (85)$ \\
(9, 0)      & $27.2022 (84)$    &             &                  & $27.339 (51)$& $27.342 (135)$ \\
\hline
\end{tabular}
\tablefoot{
\tablefoottext{a}{ Theoretical correction factor to $\Delta\nu$ scaling relation according to \citep[Figure 3]{Rodrigues2017} assuming RGB star with [M/H] = -0.38 and $M = 0.93M_{\bigodot}$. Interpolated linearly based on nearest grid points on the plot.}
\tablefoottext{A}{ Standard KEPSEISMIC light curve from MAST, filtered with an 80-day filter.
See text for details.}
\tablefoottext{B}{ EB-corrected and manually filtered version of the KASOC light curve.
See text for details.}
\tablefoottext{C}{ Standard KASOC pipeline filtered light curve.}
\tablefoottext{D}{ \citep{kallinger2019}.
URL: \url{https://github.com/tkallinger/KeplerRGpeakbagging/blob/master/ModeFiles/10001167.modes.dat}}
}
\end{table*}
\subsection{Individual mode frequency inference}\label{sec:inference-details}
\begin{figure*}
    \makebox[\textwidth][c]{\includegraphics[width=175mm, height=220mm, keepaspectratio]{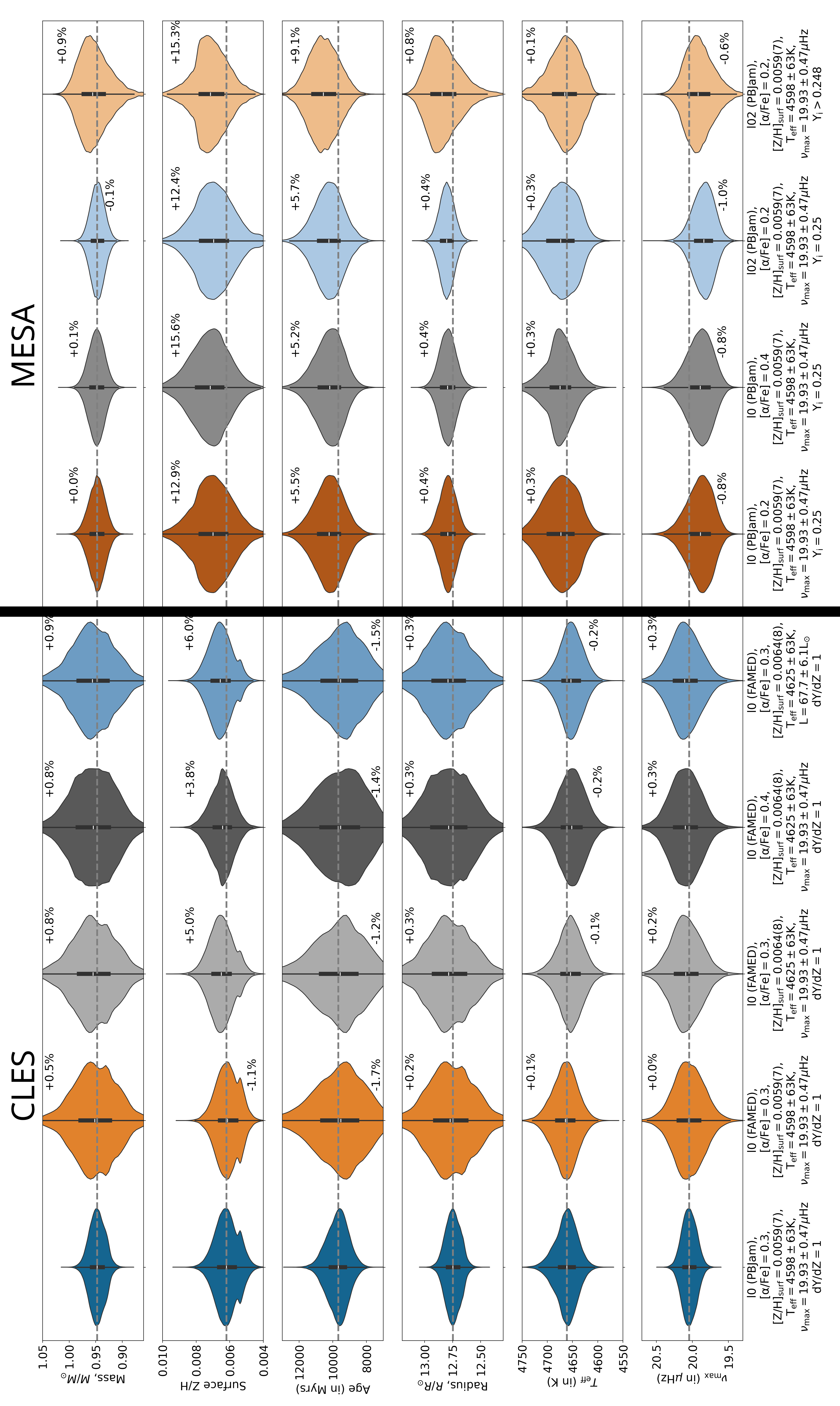}}%
    \caption{Violin plots of key stellar parameters for the 8 inferences used to define the asteroseismic measurement in Table~\ref{table:RGdata}, as well as for the helium-free inference (right-most). Left-most inference defines the reference for the median measurement and statistical uncertainty in  Table~\ref{table:RGdata}, while the maximum median difference to the remaining 7 constrained-helium inferences defines the systematic uncertainty on the quoted parameters. See the Sect.~\ref{sec:inference} and Appendix~\ref{sec:inference-details}} for details. The median parameter values of the reference fit is drawn through the whole plot (grey dashed line). The helium-free inference here has a lower bound on initial helium of 0.248 corresponding to the primordial helium fraction.
    \label{fig:violin_ast}
\end{figure*}

We perform several runs of the AIMS code, changing either to the MESA model grid ($[\rm \alpha/Fe] = 0.2$), the CL\'ES grid with $[\rm \alpha/Fe] = 0.4$, using luminosity instead of $\nu_{\rm max}$, switching between the two metallicity and effective temperature sources, or using a different set of observed radial mode frequencies (KASOC+FAMED). These fits are illustrated in Figure~\ref{fig:violin_ast}. We find that the inferred masses and radii are largely consistent across the runs, and we use this procedure to evaluate realistic systematic uncertainties on the quoted stellar parameters by taking, for each inferred parameter, the largest difference of the median between our reference run and all the aforementioned runs. The right-most run in Figure~\ref{fig:violin_ast} (MESA, helium free but constrained to $>0.248$) is not taken into consideration for systematic uncertainty estimation, but will be compared with in the next paragraph.

\begin{figure*}
     \makebox[\textwidth][c]{\includegraphics[width=175mm, height=220mm, keepaspectratio]{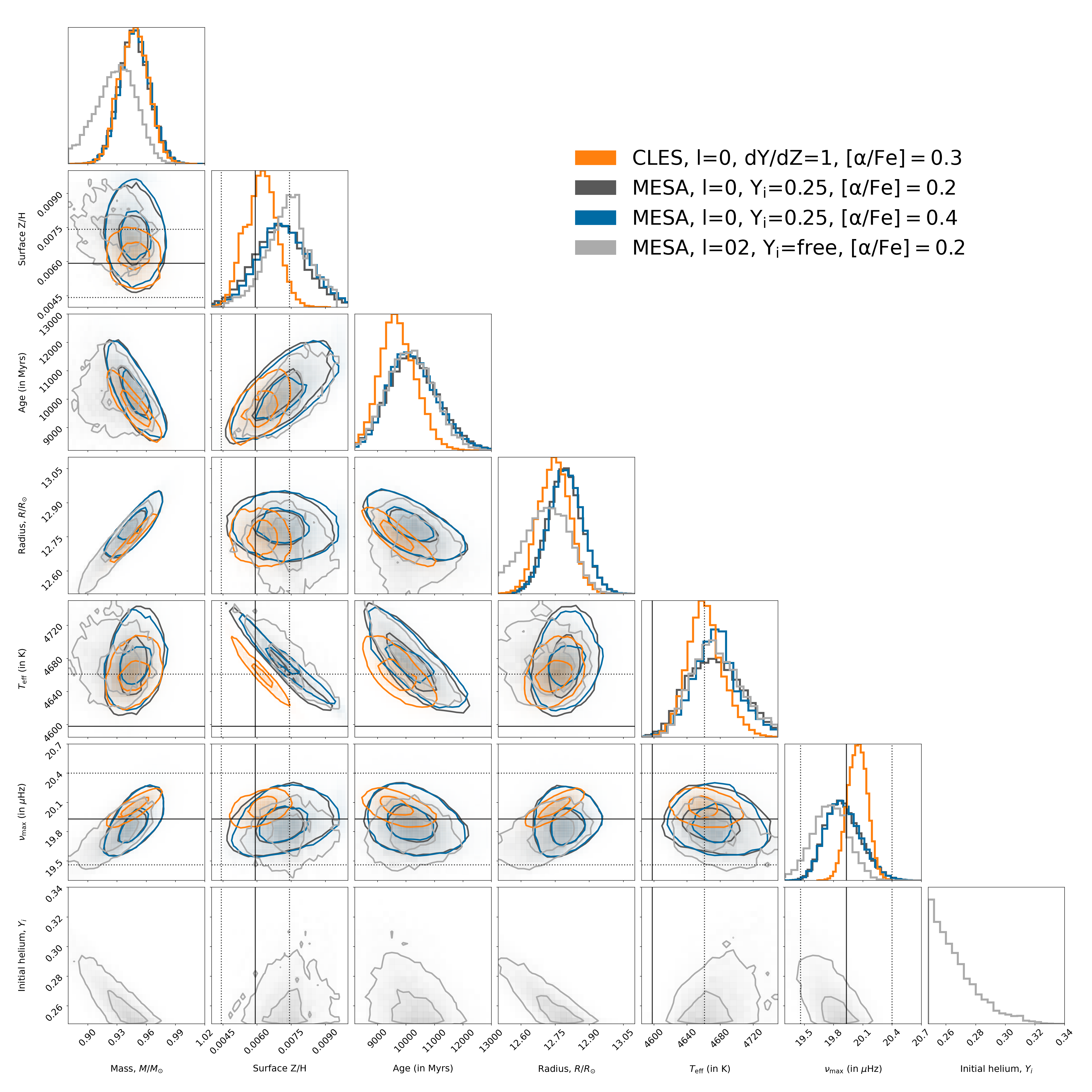}}%
    \caption{Corner plot showing a small selection of the asteroseismic inferences, all with the same Gaussian observational constraints highlighted by the full lines, $\pm$ the dotted lines indicating the $1\sigma$ level uncertainty on surface metal fraction, effective temperature, and frequency of maximum power. The grey MESA fit samples initial helium fraction as a free parameter. The lower bound on helium fraction is $0.248$. The asteroseismic frequency constraints employed were KEPSEISMIC+PBJam (see Sect.~\ref{sec:freq_numax} for description).}
    \label{fig:corner_ast}
\end{figure*}

Figure~\ref{fig:corner_ast} compares the CL\'ES reference inference with three MESA inferences, all having the same source of asteroseismic (PBJam) and photospheric constraints (APOGEE DR17). The three MESA runs are; l=0 modes and $\rm [\alpha/Fe]=0.2$, l=0 modes and $\rm [\alpha/Fe]=0.4$, and finally l=0,2 modes and $\rm [\alpha/Fe]=0.2$ without fixing the initial helium fraction or using a helium enrichment law. Most importantly, changing the grid clearly has no significant effect on the inferred mass, and the age difference between the two can be associated with a difference in temperature scale. The initial helium mass fraction is not well-constrained, yet it is compatible with the expected close to primordial value. While some correlation is seen between mass and helium, the mass agrees to $1\sigma$ with the other results.

\begin{table}
\centering
\caption{\label{table:RGdata_ast} Asteroseismic and dynamical measurements of the RG in KIC\,10001167, all using the same photospheric constraints (APOGEE DR17) unless otherwise stated. For asteroseismic scaling relations, the following stellar and solar values are used: $\nu_{\rm max}=19.784\pm0.158\mu$Hz, $\Delta\nu=2.714\pm0.017\mu$Hz, $T_{\rm eff}=4625\pm60$K, $f_{\Delta\nu}=0.95863$, $\nu_{\rm max, \odot}=3090\mu$Hz, $\Delta\nu_{\odot}=134.9\mu$Hz, $T_{\rm eff, \odot} = 5772$K.}
\begin{tabular}{lll}
\hline
\hline
Quantity & \multicolumn{2}{c}{Value} \\
\hline
Mass$_{\rm dyn} ~(M_{\odot})$                        &\multicolumn{2}{c}{$0.9337 (77)$}\\
Mass$_{\rm sis,CL\acute{E}S,l=0 (PBJam)} ~(M_{\odot})$         &\multicolumn{2}{c}{$0.947 (15)$}\\
Mass$_{\rm sis,CL\acute{E}S,l=0 (FAMED)} ~(M_{\odot})$         &\multicolumn{2}{c}{$0.954 (42)$}\\
Mass$_{\rm sis,MESA,l=0 (PBJam)} ~(M_{\odot})$        &\multicolumn{2}{c}{$0.948 (15)$}\\
Mass$_{\rm sis,MESA,l=0,2 (PBJam)} ~(M_{\odot})$        &\multicolumn{2}{c}{$0.947 (13)$}\\
Mass$_{\rm sis,MESA,l=0,2 (PBJam), Y=free} ~(M_{\odot})$  &\multicolumn{2}{c}{$0.956 (29)$}\\
Mass$_{\rm sis,scaling} ~(M_{\odot})$                       &\multicolumn{2}{c}{$1.149 (46)$}\\
Mass$_{\rm sis,scaling,corr} ~(M_{\odot})$                       &\multicolumn{2}{c}{$0.970 (39)$}\\
\hline
Radius$_{\rm dyn} ~(R_{\odot})$                  &\multicolumn{2}{c}{$13.03 (12)$}\\
Radius$_{\rm sis,CL\acute{E}S,l=0 (PBJam)} ~(R_{\odot})$                  &\multicolumn{2}{c}{$12.748 (68)$}\\
Radius$_{\rm sis,CL\acute{E}S,l=0 (FAMED)} ~(R_{\odot})$                  &\multicolumn{2}{c}{$12.77 (21)$}\\
Radius$_{\rm sis,MESA,l=0 (PBJam)} ~(R_{\odot})$                  &\multicolumn{2}{c}{$12.793 (75)$}\\
Radius$_{\rm sis,MESA,l=0,2 (PBJam)} ~(R_{\odot})$               &\multicolumn{2}{c}{$12.803 (66)$}\\
Radius$_{\rm sis,MESA,l=0,2 (PBJam), Y=free } ~(R_{\odot})$      &\multicolumn{2}{c}{$12.85 (15)$}\\
Radius$_{\rm sis,scaling} ~(R_{\odot})$                  &\multicolumn{2}{c}{$14.16 (23)$}\\
Radius$_{\rm sis,scaling,corr} ~(R_{\odot})$                  &\multicolumn{2}{c}{$13.01 (21)$}\\
\hline
$\rho_{\rm dyn} ~(10^{-3}\rho_{\odot})$                &\multicolumn{2}{c}{$0.422(11)$}\\
$\rho_{\rm sis,CL\acute{E}S,l=0 (PBJam)} ~(10^{-3}\rho_{\odot})$                &\multicolumn{2}{c}{$0.45733 (92)$}\\
$\rho_{\rm sis,CL\acute{E}S,l=0 (FAMED)} ~(10^{-3}\rho_{\odot})$                &\multicolumn{2}{c}{$0.4569 (29)$}\\
$\rho_{\rm sis,MESA,l=0 (PBJam)} ~(10^{-3}\rho_{\odot})$                &\multicolumn{2}{c}{$0.4521 (46)$}\\
$\rho_{\rm sis,MESA,l=0,2 (PBJam)} ~(10^{-3}\rho_{\odot})$              &\multicolumn{2}{c}{$0.4508 (36)$}\\
$\rho_{\rm sis,MESA,l=0,2 (PBJam), Y=free} ~(10^{-3}\rho_{\odot})$      &\multicolumn{2}{c}{$0.4507 (42)$}\\
$\rho_{\rm sis,scaling} ~(10^{-3}\rho_{\odot})$                &\multicolumn{2}{c}{$0.4048 (51)$}\\
$\rho_{\rm sis,scaling,corr} ~(10^{-3}\rho_{\odot})$                &\multicolumn{2}{c}{$0.4405 (55)$}\\
\hline
$\log g_{\rm dyn} ~(\rm dex)$                   &\multicolumn{2}{c}{$2.1786 (78)$}\\
$\log g_{\rm sis,CL\acute{E}S,l=0 (PBJam)} ~(\rm dex)$                   &\multicolumn{2}{c}{$2.2037 (22)$}\\
$\log g_{\rm sis,CL\acute{E}S,l=0 (FAMED)} ~(\rm dex)$                   &\multicolumn{2}{c}{$2.2040 (52)$}\\
$\log g_{\rm sis,MESA,l=0 (PBJam)} ~(\rm dex)$                   &\multicolumn{2}{c}{$2.2005 (40)$}\\
$\log g_{\rm sis,MESA,l=0,2 (PBJam)} ~(\rm dex)$                   &\multicolumn{2}{c}{$2.1996 (32)$}\\
$\log g_{\rm sis,MESA,l=0,2 (PBJam), Y=free} ~(\rm dex)$                   &\multicolumn{2}{c}{$2.2008 (41)$}\\
$\log g_{\rm sis,scaling} ~(\rm dex)$                 &\multicolumn{2}{c}{$2.1963 (45)$}\\
$\log g_{\rm sis,scaling,corr} ~(\rm dex)$                 &\multicolumn{2}{c}{$2.1963 (45)$}\\
\hline
$\rm age_{\rm dyn, MESA, FIES} ~(Gyr)$                       &\multicolumn{2}{c}{9.95 (70)}\\
$\rm age_{\rm dyn, CL\acute{E}S, FIES} ~(Gyr)$                       &\multicolumn{2}{c}{10.33 (48)}\\
$\rm age_{\rm dyn, CL\acute{E}S, DR17} ~(Gyr)$                       &\multicolumn{2}{c}{10.16 (47)}\\
$\rm age_{sis, CL\acute{E}S, l=0 (PBJam)} ~(Gyr)$                      &\multicolumn{2}{c}{9.68 (64)}\\
$\rm age_{sis, CL\acute{E}S, l=0 (FAMED)} ~(Gyr)$                      &\multicolumn{2}{c}{9.5 (16)}\\
$\rm age_{sis, MESA, l=0,2 (PBJam)} ~(Gyr)$                    &\multicolumn{2}{c}{10.23 (87)}\\
$\rm age_{sis, MESA, l=0,2 (FAMED), Y=free} ~(Gyr)$            &\multicolumn{2}{c}{10.56 (88)}\\
$\rm age_{sis, MESA, l=0 (PBJam)} ~(Gyr)$                      &\multicolumn{2}{c}{10.21 (90)}\\

\hline
\end{tabular}
\end{table}
\subsection{Asteroseismic mean density inversion}\label{sec:inversion}
The consistency of the results provided from the fit of the radial modes was tested using a mean density inversion following \citet{buldgen2019}.
To this end, we used the two sets of radial modes (KEPSEISMIC+FAMED, and KEPSEISMIC+PBJam) with respectively 5 and 6 individual modes and two reference models computed with CL\'ES, and optimized with AIMS to reproduce the individual radial frequencies.
To fully test the precision of the inversion, we also tested various surface corrections, namely that of \citet{ballgizon2014}, that of \citet{Sonoi2015} and leaving no surface correction.
The average of all these results provides a value of $6.39\pm 0.04\times 10^{-4}$ $\rm g/cm^{3}$ with two distinct families of solutions, one centered around $6.43\times 10^{-4}$ $g/cm^{3}$ when surface corrections are included, one centered around $6.35\times 10^{-4}$ $\rm g/cm^{3}$ without.
This behaviour is typical of mean density inversions, where systematics will heavily dominate the total uncertainty budget.
Ultimately, the inversion confirms the AIMS modelling which in retrospect is not surprising as AIMS succeeded in fitting the individual radial modes as primary data in the forward modelling procedure.
In this particular case, no specific disagreement in mean density is observed.
\subsection{Scaling relations}\label{sec:scaling-relations}
The asteroseismic scaling relations for mass and radius are given below \citep[e.g.][and references therein]{Sharma2016}:

\begin{eqnarray}\label{eq:03}
\frac{M}{\mathrm{M}_\odot} & \simeq & \left(\frac{\nu _{\mathrm{max}}}{f_{\nu _{\mathrm{max}}}\nu _{\mathrm{max,}\odot}}\right)^3 \left(\frac{\Delta \nu}{f_{\Delta \nu}\Delta \nu _{\odot}}\right)^{-4} \left(\frac{T_{\mathrm{eff}}}{T_{\mathrm{eff,}\odot}}\right)^{3/2},\\
\label{eq:04}
\frac{R}{\mathrm{R}_\odot} & \simeq & \left(\frac{\nu _{\mathrm{max}}}{f_{\nu _{\mathrm{max}}}\nu _{\mathrm{max,}\odot}}\right) \left(\frac{\Delta \nu}{f_{\Delta \nu}\Delta \nu _{\odot}}\right)^{-2} \left(\frac{T_{\mathrm{eff}}}{T_{\mathrm{eff,}\odot}}\right)^{1/2}. 
\end{eqnarray}
$\Delta\nu$ is the large frequency spacing in the asteroseismic power spectrum, and $\nu_{\rm max}$ is its frequency of maximum power.
We use the solar reference values of $\Delta \nu _{\odot} = 134.9 \mu$Hz and $\nu _{\rm max,\odot} = 3090 \mu$Hz following \citet{Handberg2017}, and $T_{\rm eff, \odot} = 5772$K from IAU 2015 B3 \citep{IAU2015b3}.

The correction factor $f_{\Delta\nu}$ is a stellar modelling derived correction to account for the difference in internal structure between the red giant and the Sun.
When referring to the scaling relations in this work, we utilize the radial-mode-only $\Delta\nu_0=2.714 \pm 0.017 \mu$Hz, $\nu_{\rm max} = 19.784 \pm 0.158 \mu$Hz, $f_{\Delta\nu}$ from \citet{Rodrigues2017}, and $f_{\nu_{\rm max}} = 1$.

Asteroseismic scaling relation measurements are provided in Table~\ref{table:RGdata_ast}. The corrected relations show an overall agreement with the individual frequency inference, but a $3.6\%$ difference in average density ($3\sigma$), which is much smaller than the disagreement with eclipsing binary analysis. The mass obtained with the scaling relation is $0.970\pm 0.036 \rm M_{\odot}$, which is $1\sigma$ consistent with the individual frequency inferences mainly due to the comparatively large statistical uncertainty. As these scalings are very sensitive to the assumptions surrounding measurement of $\nu_{\rm max}$, $\Delta\nu$, and the model-inferred $\Delta\nu$ correction, they can be systematically much more uncertain than the individual frequency inference. This is the reason that this work focuses on comparisons with the latter method for the purpose of ascertaining the accuracy of asteroseismic mass measurements.
\section{Evolutionary stage of the red giant}\label{sec:evol_state}
\begin{figure}
    \centering
    \includegraphics[width=\columnwidth]{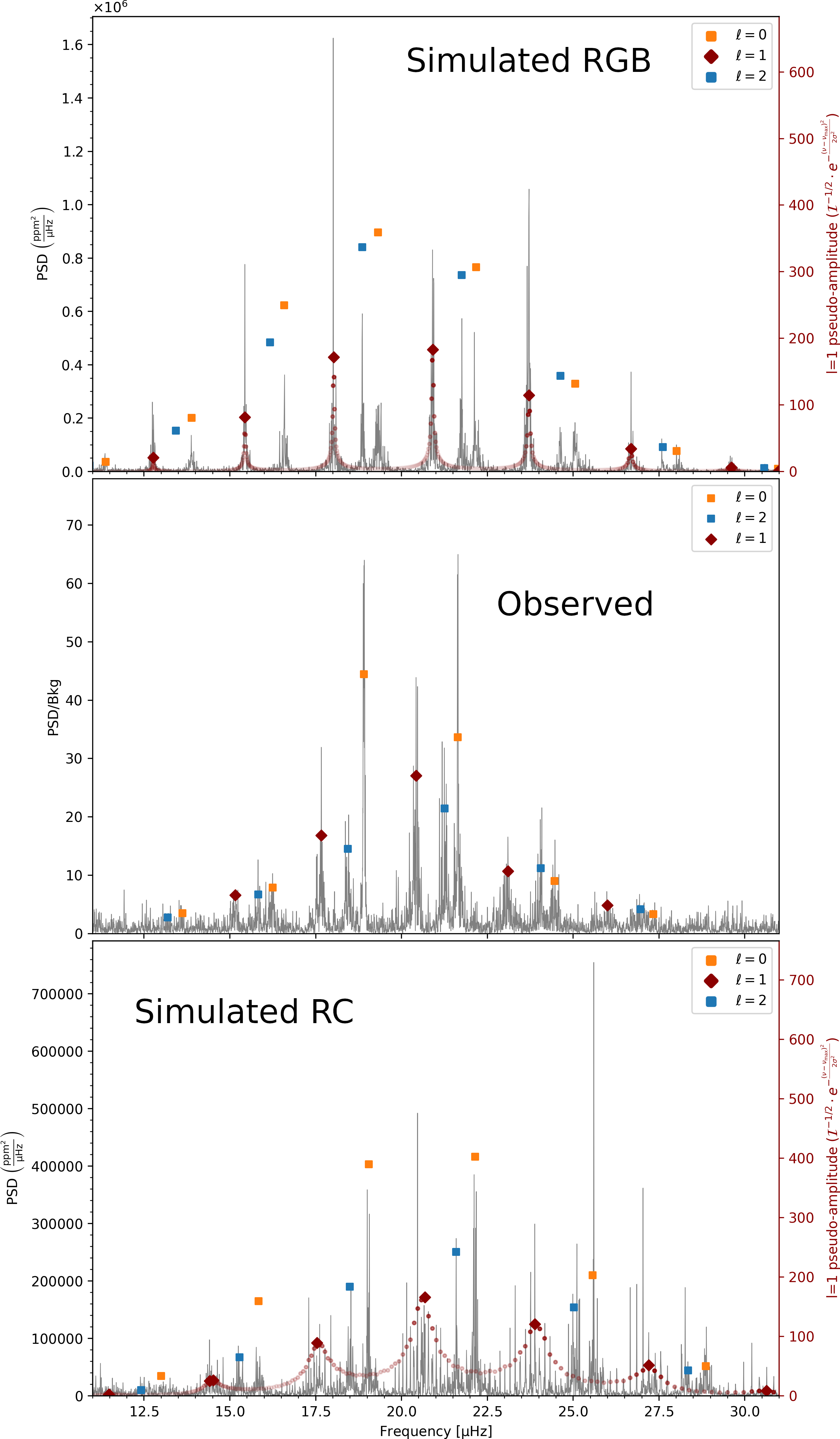}
    \caption{Top panel: Simulated power spectral density (PSD) for RGB star. Middle panel: Observed PSD, divided by granulation background. Bottom panel: Simulated PSD for a late-clump/early AGB star. For the simulated PSD, the mixed $\ell$=1 modes are highlighted, with an amplitude proxy calculated from the mode inertia and a Gaussian envelope.}
    \label{fig:sim_ast}
\end{figure}

We attempted, with well-established methods \citep{Mosser2012}, to measure an asymptotic period spacing from the dipole modes of the RG, and found that we were unable to clearly resolve any mixed dipole modes. If the RG had been core-Helium burning, this should have been straightforward as the period spacing would take on large values causing very well-separated mixed modes \citep[see e.g.][]{Mosser2012,Mosser2014}. This is what we furthermore demonstrate with synthetic asteroseismic spectra in Fig.~\ref{fig:sim_ast}, and we detail their calculation below.

We also attempted an asteroseismic inference assuming that the RG is past the RGB, and fitted the radial modes only. The best-fit obtained this way is found at the very end of the red clump (RC), while a small fraction of He is still being burned in the core.
With this fit, the surface correction coefficients take on unphysical values, giving it the appearance of a linear frequency shift rather than an asymptotic behaviour towards lower frequencies.
The temperature recovered with this fit is more than 200K higher than observed.

Fig.~\ref{fig:sim_ast} compares the observed power spectrum with simulated asteroseismic power spectral densities calculated with the code {\scshape AADG v3.0.2} \citep[AsteroFLAG Artificial Dataset Generator,][]{Ball2018}, one for a late-RC/early-Asymptotic Giant Branch (AGB) similar to the radial mode only post-RGB solution described above, and one for the RGB.
These power spectral densities simulate 4-year-long {\it Kepler} observations of a red-giant star with $\rm M = 0.95 \, M_\odot, \, [Fe/H] = -0.75$ and $\rm [\alpha/ Fe] = 0.4$, with $\nu_{\rm max}$ and $\Delta\nu$ approximately equivalent to the observed values; we adopted the values of the radial mode linewidths observed in large samples of RGB and early-AGB stars \citep{Vrard2018,Dreau2021} in these simulations \citep[i.e. similarly to][]{Matteuzzi2023}.
These clearly show that the RGB phase is the most credible phase for KIC\,10001167, since the observed mixed-mode behavior of the dipole ($\ell$=1) modes is only compatible with the RGB simulation.

\subsection{Further evolutionary considerations from population membership and binary orbit}

The kinematics, metallicity, and $\alpha$-enhancement points towards KIC\,10001167 being a member of the old, thick disk population of the Milky Way.
If the RG has evolved past the RGB through single-star evolution, it would have gone through the He flash and lost a significant amount of mass. Its initial mass would be much higher, and therefore the system should then be younger than what is compatible with the rest of the old disk population.

The RG currently has a radius of about $\sim 10\%$ the size of the semi-major axis ($13.03 \rm R_{\odot}$ vs. $124.17 \rm R_{\odot}$). From stellar models, the red giant branch tip radius of a star with this mass is expected to be $\sim \rm 120 \rm R_{\odot}$. Given the current orbital separation and eccentricity, if the star had evolved past the tip its companion must have entered the convective envelope at that point during its evolution. It is unlikely that the binary would have survived this, further demonstrating that the RG must be on the RGB.
\section{Influence of tides on the red giant}\label{sec:tides}
As shown in Fig.~\ref{fig:phoebe_10001167}, the light curve has a clear tidal deformation signal from the RG in KIC\,1001167.
However, this still amounts to only a small level of deformation from spherical.
When assuming the stellar shape as a bi-axial ellipsoid, the approximated oblateness $(a-b)/a$ is less than 0.002.
This means that, for all intents and purposes besides fitting the filtered light curve, the star can be assumed spherical.

\citep{beck2023} suggests that radius expansion, caused by tidal interaction with a companion, could lead to overestimation of the seismically inferred mass and radius.
So far, this claim does not seem to have been tested.
The assumption underlying this is that there should be a natural change in internal structure profile due to interaction with the companion, in a way that perturbs the seismic profile significantly from what can be predicted using traditional stellar modelling.

KIC\,10001167 is not in a regime where detailed investigation is relevant.
The level of tidal apsidal motion is expected to be very small at its current evolutionary state, and we have not been able to fit a significant linear apsidal motion to the light curve and radial velocities.
Angular momentum transfer between the RG internals and the binary orbit is therefore currently below detectable levels, indicating that the tidal displacement of matter is still low despite showing significant flux variation.
Additionally, we have performed two tests using representative stellar evolutionary tracks calculated with MESA. Both have $\rm [Fe/H]=-0.75$, $\rm [\alpha/Fe]=0.4$, $Y=0.25$, with either initial mass $0.90 \rm M_{\odot}$ or $0.95 \rm M_{\odot}$.
We calculated the circularization function from \citet[eq. 7]{beck2023} by integrating numerically from zero-age main sequence up to the age where the radius of the star surpasses $13.03 \rm R_{\odot}$.
With this, we obtain a rate of eccentricity reduction of $\epsilon_r = -0.37$ for initial mass $0.90 \rm M_{\odot}$ and $\epsilon_r = -0.42$ for $0.95 \rm M_{\odot}$ (using \citep[eq. 5, eq. 6]{beck2023}). This is significantly below the critical $\epsilon_{\rm crit} = 0.478$ which is suggested by \citet{verbunt1995} to separate systems with strong and weak tides.
This demonstrates that tidal history is not a likely explanation for the differences we see between asteroseismology and eclipsing binary measurements of KIC\,10001167.
\section{Mass-loss on the red giant branch}\label{sec:mass-loss}
Our stellar models do not include mass loss during the RGB, which does not impact our main findings, meaning our tests on the mass recovery of asteroseismology. If mass loss is included in stellar models, it is typically through simple analytic mass loss prescriptions \citep[e.g.][]{Reimers1975}. In reality, the exact occurrence of mass loss during the RGB has not been observationally established, rather, integrated mass loss from RGB to red clump is better quantified \citep[see e.g.][]{Brogaard2024}. If we assume the \citet{Reimers1975} prescription, the integrated RGB--RC mass loss measurements of \citet{Brogaard2024} would indicate $\eta \sim 0.4$ for a star with the properties of KIC\,10001167. With $\eta=0.2, 0.4, 0.6$, we can predict that the RG should then have lost $0.6, 1.2, 1.8$\% of its mass, respectively. Ignoring it demonstrates a potential systematic uncertainty on the age of $2, 4, 6$\% for this star depending on adopted $\eta$. All these estimates are roughly equivalent to or below our adopted systematic uncertainty. We stress again that there is currently no observational indication of significant mass loss at luminosities comparable to that of this object.
\section{All radial velocity measurements}
\begin{table*}
\centering
\addtolength{\leftskip}{-3cm}
\addtolength{\rightskip}{-3cm}
\begin{tabular}{lllllllllllllll}
\hline
BJD-X &
  S/N &
  W &
  $\rm rv_{RG, BF}$ &
  $\rm rv_{MS, BF}$ &
  $\rm \sigma_{RG, BF}$ &
  $\rm \sigma_{MS, BF}$ &
  $\rm rv_{CCF}$ &
  $\rm \sigma_{CCF}$ &
  $\rm FWHM$ &
  $\rm con$ &
  $\rm bis$ &
  $\rm MS$ \\ \hline
8385.4869  & 49 & 0.7 & -92.749  & -113.420 & 0.095 & 0.515 & -92.266  & 0.091  & 9.364  & 21.138 & 0.1406 & 1     \\
8386.4297  & 41 & 0.6 & -93.561  & -113.090 & 0.095 & 0.750 & -93.080  & 0.091   & 9.394  & 21.022 & 0.0954 & 1     \\
8386.4521  & 49 & 0.7 & -93.618  & -113.077 & 0.096 & 0.482 & -93.144  & 0.091   & 9.425  & 21.158   & 0.1590 & 0    \\
8392.4546  & 60 & 0 & -99.187  & ---      & 0.096 & ---   & -98.731   & 0.091  & 9.302  & 21.460 & 0.1362  & 0    \\
8393.4032  & 58 & 0 & -100.132 & ---      & 0.096 & ---   & -99.642  & 0.091  & 9.348  & 21.166 & 0.1430  & 0    \\
8396.3671  & 46 & 0 & -103.213 & ---      & 0.098 & ---   & -102.737 & 0.091   & 9.528  & 21.332 & 0.1555 & 0    \\
8407.4649  & 52 & 0 & -116.255 & -88.007  & 0.097 & 0.485 & -115.788  & 0.091  & 9.242  & 20.938 & 0.1507  & 1     \\
8426.3591  & 54 & 0.7 & -131.251 & -71.169  & 0.095 & 0.145 & -130.763 & 0.091   & 9.388  & 20.951 & 0.1358  & 1     \\
8427.3655  & 57 & 0.7 & -131.181 & -70.365  & 0.097 & 0.549 & -130.704  & 0.090  & 9.367  & 20.965 & 0.1043 & 1     \\
8428.4015  & 41 & 0.7 & -131.074 & -71.323  & 0.097 & 0.329 & -130.569  & 0.091   & 9.533  & 20.842 & 0.1139 & 1     \\
8429.3965  & 45 & 0.7 & -130.792 & -72.091  & 0.096 & 0.373 & -130.307   & 0.091  & 9.346  & 20.985 & 0.1244 & 1     \\
8437.3641  & 53 & 0.7 & -124.212 & -78.969  & 0.096 & 0.270 & -123.739  & 0.091  & 9.408  & 20.919  & 0.1547 & 1     \\
8446.3035  & 49 & 0 & -110.724 & -94.231  & 0.096 & 0.466 & -110.236 & 0.091  & 9.598  & 20.861 & 0.1646 & 0    \\
8454.3115  & 54 & 0 & -98.399  & ---      & 0.096 & ---   & -97.943   & 0.091  & 9.303  & 21.063 & 0.1308 & 0    \\
8458.3207  & 19 & 0.7 & -93.362  & -113.348 & 0.098 & 0.932 & -92.868  & 0.098   & 9.247  & 20.366  & 0.2036 & 0    \\
8461.3147  & 31 & 0.7 & -90.055  & -117.316 & 0.094 & 0.887 & -89.628  & 0.092  & 9.300  & 20.792 & 0.1831 & 1     \\
9036.5826  & 70 & 1 & -127.251 & -75.462  & 0.094 & 0.228 & -126.791  & 0.091  & 9.225  & 21.034 & 0.1368  & 1     \\
9037.6167  & 49 & 1 & -126.145 & -77.410  & 0.094 & 0.694 & -125.718 & 0.091  & 9.446  & 21.0786 & 0.1349 & 1     \\
9038.5733  & 62 & 1 & -125.039 & -78.325  & 0.095 & 0.462 & -124.585 & 0.091  & 9.362  & 21.178 & 0.1502 & 1     \\
9039.6590  & 64 & 1 & -123.758 & -79.740  & 0.094 & 0.361 & -123.336 & 0.091  & 9.358  & 21.191 & 0.1431 & 1     \\
9047.5846  & 11 & 0 & -111.847 & -91.784  & 0.113 & 1.432 & -111.314 & 0.129  & 10.201 & 20.820 & 0.1699 & 0    \\
9072.5926  & 84 & 1 & -83.571  & -124.487 & 0.096 & 0.560 & -83.100  & 0.091  & 9.193  & 20.897  & 0.1915 & 1     \\
9077.5191  & 85 & 1 & -81.939  & -125.561 & 0.096 & 0.248 & -81.505  & 0.091  & 9.088  & 21.135 & 0.1724 & 1     \\
9078.5015  & 88 & 1 & -81.747  & -126.538 & 0.096 & 0.430 & -81.312  & 0.090  & 9.242  & 21.180 & 0.1424 & 1     \\
9079.5529  & 78 & 1 & -81.597  & -126.711 & 0.097 & 0.449 & -81.155  & 0.090  & 9.254  & 21.181  & 0.1573 & 1     \\
9082.3838  & 86 & 1 & -81.402  & -127.040 & 0.095 & 0.349 & -80.976  & 0.090  & 9.106   & 21.109  & 0.1405  & 1     \\
9380.5590  & 63 & 0.9 & -127.003 & -76.544  & 0.095 & 0.332 & -126.538 & 0.091  & 9.248  & 21.081  & 0.1776  & 1     \\
9429.4490  & 44 & 0.7 & -86.068  & -121.504 & 0.094 & 0.608 & -85.594  & 0.091  & 9.387  & 20.671 & 0.1456 & 1     \\
9433.5892  & 54 & 0.7 & -83.692  & -124.280 & 0.096 & 0.461 & -83.250  & 0.091  & 9.392  & 20.690 & 0.1399 & 1     \\
9446.5612  & 78 & 0.9 & -81.757  & -126.571 & 0.096 & 0.554 & -81.331  & 0.090  & 9.185  & 20.639  & 0.1370 & 1     \\
9863.4595  & 56 & 0.7 & -128.353 & -74.807  & 0.096 & 0.258 & -127.918 & 0.091  & 9.079  & 20.677 & 0.1590 & 1     \\
9989.7557  & 27 & 0 & -131.301 & -70.671  & 0.098 & 0.792 & -130.886 & 0.093  & 9.370  & 20.538 & 0.2043  & 0    \\
9996.7724  & 36 & 0 & -129.837 & -72.915  & 0.095 & 0.355 & -129.376 & 0.091  & 9.194   & 20.542 & 0.1681 & 1     \\
10002.7624 & 53 & 0 & -123.889 & -79.214  & 0.094 & 0.328 & -123.474 & 0.091  & 9.320  & 20.646 & 0.1718 & 1     \\
10007.7233 & 46 & 0 & -116.626 & -87.391  & 0.095 & 0.733 & -116.187 & 0.091  & 9.217  & 20.727 & 0.1458 & 1     \\
10015.6930 & 41 & 0 & -103.809 & ---      & 0.096 & ---   & -103.343  & 0.091  & 9.451  & 21.187  & 0.1900 & 0    \\
10021.7254 & 53 & 0 & -95.333  & -111.835 & 0.096 & 0.375 & -94.892  & 0.091  & 9.167  & 21.100 & 0.1919 & 0    \\
10032.7096 & 23 & 0 & -84.995  & ---      & 0.098 & ---   & -84.521  & 0.094  & 9.616  & 20.694 & 0.2135 & 0    \\
10033.6893 & 16 & 0 & -84.393  & ---      & 0.105 & ---   & -83.995   & 0.099  & 8.974  & 20.751 & 0.1825 & 0    \\
10033.7385 & 21 & 0 & -84.366  & ---      & 0.100 & ---   & -83.888  & 0.093  & 9.141   & 20.875 & 0.0977 & 1     \\
10036.6912 & 37 & 0 & -82.951  & ---      & 0.097 & ---   & -82.495   & 0.091  & 9.439   & 20.765   & 0.1516  & 1     \\
10042.6514 & 63 & 0 & -81.460  & -126.517 & 0.095 & 0.344 & -80.995  & 0.091  & 9.292  & 20.677 & 0.1585 & 1     \\
10047.6683 & 33 & 0 & -81.385  & -127.532 & 0.097 & 0.537 & -80.943  & 0.091  & 9.714   & 20.521 & 0.1500 & 1     \\
10061.5616 & 48 & 0 & -86.226  & -121.694 & 0.094 & 0.288 & -85.799  & 0.097 & 9.787  & 20.617 & 0.1626  & 1     \\
10069.6970 & 58 & 0 & -91.893  & -115.821 & 0.096 & 0.623 & -91.457  & 0.098 & 9.240  & 20.694 & 0.1819 & 1    \\\hline
\end{tabular}
\caption{\label{table:rvs} Table detailing the spectroscopic radial velocity follow-up observations for KIC\,10001167. Columns, and their units, are as follows: \emph{BJD-X} is the barycentric julian date - 2450000. \emph{S/N} is estimated Signal/Noise ratio of the full frame images at $\sim$5880Å. \emph{W} is the weight applied to each spectrum when performing the spectral separation. $\rm rv_{RG, BF}$ and $\rm rv_{MS, BF}$ [km/s] are the radial velocities of the giant and main sequence star from the main broadening function analysis. $\rm \sigma_{RG, BF}$ and $\rm \sigma_{MS, BF}$ [km/s] are the corresponding RV uncertainties for the giant and main sequence star used in the eclipsing binary analysis. Similar results are reported for the independent CCF analysis of the RG. For that analysis, we also carried out full-width at half maximum (FWHM) [km/s], line contrast (con), and bisector-span (bis) measurements. Last, the spectra included for MS component fit with the CCF method is reported in ($\rm MS$).}
\end{table*}
\end{appendix}
\end{document}